\documentclass[fleqn]{article}
\usepackage{a4wide}
\usepackage[utf8]{inputenc}
\usepackage{amsmath}
\usepackage{titlesec}
\usepackage[colorlinks = true,
            linkcolor = blue,
            urlcolor  = blue,
            citecolor = blue,
            anchorcolor = blue]{hyperref}
\titlespacing*{\section}
{0pt}{0.4ex plus 0.1ex minus .1ex}{0.3ex plus .2ex}
\titlespacing*{\subsection}
{0pt}{0.3ex plus 0.1ex minus .1ex}{0.3ex plus .2ex}
\newcommand\Mark[1]{\textsuperscript#1}

\usepackage{xcite}
\usepackage{xr}
\makeatletter
\newcommand*{\addFileDependency}[1]{
  \typeout{(#1)}
  \@addtofilelist{#1}
  \IfFileExists{#1}{}{\typeout{No file #1.}}
}
\makeatother

\title{A sequential Monte Carlo approach to estimate a time varying reproduction number in infectious disease models: the Covid-19 case.}

\author{Geir Storvik\Mark{1} \and Alfonso Diz-Lois Palomares\Mark{2} \and Solveig Engebretsen\Mark{3} \and Gunnar Øyvind Isaksson Rø\Mark{2}  \and Kenth Engø-Monsen\Mark{4} \and Anja Bråthen Kristoffersen \Mark{2} \and Birgitte Freiesleben de Blasio\Mark{2}\Mark{,}\Mark{5} \and Arnoldo Frigessi\Mark{5}\Mark{,}\Mark{6}}

\date{\today}

\usepackage{amsmath}
\usepackage[round]{natbib}
\usepackage{bm}
\usepackage{graphicx}
\usepackage{algorithm}
\usepackage{algpseudocode}
\usepackage{amssymb}
\usepackage{mathtools}

\usepackage{tikz}
\usepackage{tikz-cd}

\colorlet{myred}{black}

\begin{document}
\maketitle

\noindent \Mark{1}\textit{Department of Mathematics, University of Oslo}\\
\Mark{2}\textit{Department of Method Development and Analytics, Norwegian Institute of Public Health}\\
\Mark{3}\textit{Norwegian Computing Center}\\
\Mark{4}\textit{Telenor Research, Fornebu, Norway}\\
\Mark{5}\textit{Oslo Centre for Biostatistics and Epidemiology, University of Oslo}\\
\Mark{6}\textit{Oslo Centre for Biostatistics and Epidemiology, Oslo University Hospital}\\
\begin{abstract}

\end{abstract}
During the first {\color{myred}eighteen} months, the Covid-19 pandemic has required most countries to implement complex sequences of  {\color{myred}non-pharmaceutical} interventions, with the aim of controlling the transmission of the virus in the population. To be able to take rapid decisions, a detailed understanding of the current situation is necessary. Estimates of time-varying, instantaneous reproduction numbers represent a way to quantify the viral transmission in real time. They are often defined through a  mathematical compartmental model of the epidemic, like a stochastic SEIR model, whose parameters must be estimated from multiple  time series of epidemiological data. Because of very high dimensional parameter spaces {\color{myred}(partly due to the stochasticity in the spread models)} and incomplete and delayed data, inference is very challenging. We propose a state space formalisation of the model and a sequential Monte Carlo approach which allow to estimate a daily-varying reproduction number for the Covid-19 epidemic in Norway with sufficient precision, on the basis of daily hospitalisation and positive test incidences. The method is in  {\color{myred}regular} use in Norway and is a powerful instrument for epidemic monitoring and management. 

\vskip 5mm
\section{Introduction}

We propose a dynamic approach for the estimation of a time-varying or instantaneous reproduction number for a mathematical infectious disease spread model. We apply our method to the Covid-19 pandemic in Norway. Like in most other countries, the pandemic has been tackled with a combination of {\color{myred}non-pharmaceutical} interventions, from social distancing to partial lock-down, imposed or advised at various time points.
{\color{myred}Various viral variants with different characteristics have been competing in the population. Vaccination has also been gradually introduced. 
As a consequence of these changes, which can emerge both abruptly and smoothly, the reproduction number varies.} Instantaneous estimates of reproduction numbers are useful for situational awareness. Being able to estimate such changes rapidly is important in guiding decision makers in future policy planning. 

A reproduction number is precisely defined within a mathematical model of transmission. A large class of models, which has been shown to be very appropriate, is the so called SEIR model (S=susceptible, E=exposed, I=infected, R=recovered). {\color{myred} In particular, \emph{stochastic} compartmental models are preferable when there are relatively low number of infections~\citep[chapter 6]{keeling2011modeling}}.
SEIR models are parametrised so that a meaningful estimate of the reproduction number (basic and effective) can be derived. Such compartmental models are based on many latent variables, in particular the number of individuals in each compartment  {\color{myred}within each region} at every time point, and depend on many epidemiological parameters, including the transmission strengths which are a key component of reproduction numbers. All unknowns must be estimated from data, which in a realistic situation are scarce and incomplete. For this reason, inference is in general very difficult, because of a high dimensional parameter space, rather flat likelihoods or posterior distributions and, often, weak identifiability~\citep{de2015four}.

Moreover, data which carry information about the transmissibility of the virus appear with an inevitable delay. In this paper we use two data sources, the daily number of hospitalised Covid-19 patients and the daily number of positive laboratory-confirmed RNA test cases. Both these data sources carry information about the transmission of the virus in the society, but with a random time delay from transmission.
Therefore, the uncertainty of the estimates of a daily reproduction number increases in the last period of data. All this is particularly challenging during an emerging epidemic. 
For these reasons, instantaneous reproduction numbers are rarely assumed in SEIR models. In this paper we propose and test a sequential Monte Carlo (SMC) approach to inference which allows the efficient estimation of instantaneous reproduction numbers for the Covid-19 pandemic from actual data.

While our SMC approach is generic, we implement it {\color{myred}on top of} a stochastic SEIR model which we developed for the Covid-19 epidemic~\citep{FHImodelPaper2021} and which is in {\color{myred}regular} use by the Norwegian health authorities. This model assumes a spatial scale resolved on county level, and uses mobile phone mobility data for the geographical spread of the virus in temporal steps of six hours. 
In~\citet{FHImodelPaper2021}, the  transmissibility parameter, which represents the probability of transmission upon a contact times the contact rate in the
population, is assumed to be constant in time, and is only changed at designed change-points. Inference in~\citet{FHImodelPaper2021} is performed by a version of Approximate Bayesian Computation (ABC). MCMC convergence was essentially impossible to reach, because of the difficulty to design parameter perturbations which would lead, through the stochastic SEIR, to minor and controlled changes of the posterior distribution. It is very difficult to use ABC when the number of parameters is large, as is the case when including daily varying reproduction numbers. 

In this paper we propose to perform Bayesian inference in combination with SMC. Static parameters related to the dynamic process for the reproduction number are estimated sequentially through SMC using sufficient statistics \citep{fearnhead2002markov,storvik2002particle}.
Application of SMC methods
is challenging because the latent processes are of high dimensions, the SEIR model is only available as a computer algorithm, and data are very limited. 
A further very important practical aspect when working in real time during the pandemic, comes from the continuous need to improve the model, to change epidemiological assumptions, to use different data registries, and to improve the computational efficiency of algorithms. For these reasons, it has been very important to develop an SMC which is very  flexible, so that changes in model specifications are easy to implement. This paper shows how a careful design of combination of  model and algorithm reaches these aims and leads to a good fit to both data sources. We produce an analysis of the Norwegian pandemic, quantifying how interventions
impacted the transmissibility and showing how our estimated instantaneous reproduction numbers are capturing changes rapidly enough. Based on the dynamic model, we perform future predictions of the situation for the next three weeks. We discuss the quantification of uncertainty in forecasts, which is of paramount importance for decision making.

There is an important literature on time varying reproduction numbers applied in various epidemics, see for example \citet{cauchemez2006estimating,viboud2018rapidd}.
A Bayesian framework for estimating time-varying reproduction numbers was proposed by \citet{cauchemez2006estimating}, and applied to the SARS epidemic. 
In~\citet{cori2013new}
a time varying reproduction number is defined as the ratio of the number $C(t)$ of infected in day $t$  over $\sum_k w(k) C(T-k)$, where $w(\cdot)$ is the distribution of the generation time of Covid-19, which is often set to be a gamma distribution with mean and standard deviation estimated from specific studies~\citep{ferretti2020quantifying}.
There is a very successful R-package implementing this Bayesian method, called \texttt{EpiEstim}~\citep{Cori2021}. {\color{myred}We include a comparison between \texttt{EpiEstim} and our approach based on only one of the data sources.}

Several papers have applied MCMC algorithms for estimation of parameters in compartmental models~\citep[e.g][]{gibson1998estimating,o1999bayesian,o2000analyses}.
The recent paper 
 \citet{birrell2020efficient} studies various SMC approaches in a SEIR model for influenza. It demonstrates the superiority of SMC methodology compared to MCMC for such dynamical models. The paper is also useful as a reference to SMC in epidemic modelling.  Our work shares many similarities to this approach, including the use of several sources of data. A difference is our use of a dynamic model for the reproduction numbers. Also, a stochastic delay between infection and observation time is included in our setting. Another general inference framework is implemented in the R-package \texttt{pomp}~\citep{king2016}. The package contains multiple different implementations of estimation procedures, including Sequential Monte Carlo, for inference for partially observed Markov process models. There are several examples of applications of the package to epidemic inference, see for example~\citep{king2016,stocks2020model,blackwood2013deciphering}.  
 
 The outline of the rest of the paper is as follows. In Section~\ref{sec:data} the context and the data are described. Section~\ref{sec:model} describes the full model, formulated as a state space model. In Section~\ref{sec:smc} we discuss how SMC algorithms can be applied for the inferential problem, including estimation of several static parameters. {\color{myred}A simulation study}  and experimental results are reported in Section~\ref{sec:res} specifically for the Norwegian Covid-19 pandemic. Additional results, including sensitivity analysis, are collected in the supplementary material. We conclude the paper by a summary and discussion.
 In the supplementary material details about experimental settings, algorithmic specifications. {\color{myred}Data and code availability} are available on a GitHub repository.
 
\vskip 4mm
\section{Context and data}\label{sec:data}

We start by setting the scene of the inferential task. 
The core is an existing model of the epidemic which has as input a set of parameters and variables, including daily reproduction numbers, and  as output a series of time series of infection incidence. In our case, the model is a stochastic compartmental SEIR-type model that produces numbers of susceptible, exposed, pre-symptomatic, symptomatic and asymptomatic infectious and recovered at every time point. We also keep track of the disease incidence. We use two data time series to inform the SEIR model: the daily number of new hospital admissions of Covid-19 patients, and the daily number of laboratory-confirmed positive PCR tested cases.
In order to exploit these data, we furthermore model the process of hospitalisation and testing of the SEIR output, in particular of the daily incidence of infected. The inferential task is to make inference on the input parameters. 

The hospitalisation data contain admission to all hospitals in Norway and of all patients who are diagnosed with Covid-19 as the main cause. Admission on a certain day informs us of a transmission event that has occurred some days before. This time gap can differ between individuals. We make several assumptions on various time lags, as specified in supplementary material. For example, the number of days between symptom onset and admission to hospital is estimated to be negative binomial distributed with parameters estimated 
in a separate study of the Norwegian Covid-19 registry \citep{whittaker2021trajectories}. On average, for a patient being hospitalised, the time gap between infection and hospitalisation is estimated to be approximately 14 days. For concreteness, in this paper we assumed the distributions of the various time lags to be given. 

The second data set is the time series of daily number of positive PCR tests. 
Again, there is a time gap between onset of symptoms and testing, which we estimate through a fixed distribution of delay with mean about 4 days.
The reason is that it is important for inference that the two data sets are as aligned as possible. 

{\color{myred}We use two additional data sets, which enter the SEIR model as input variables: the daily number of positive cases who have been tested in Norway but infected outside of Norway, so called imported cases; and the total number of PCR tests made in Norway, as a surrogate of the effort made to detect positive cases.}

We start with the population of Norway, distributed according to the national census in the eleven counties (see \href{https://www.ssb.no/statbank/table/12871/}{www.ssb.no/statbank/table/12871/}). 
Like in~\citet{FHImodelPaper2021}, we seed our model continuously with positive cases imported from abroad on the day of recorded symptom onset or, if not available, when detected by testing. 
Imported cases that are hospitalised are not counted in the time series of hospitalisations, because they do not inform the model about the transmissibility of the virus in Norway. Because not all imported cases are likely to be discovered, we assume that each imported case stands for an unknown number of further undetected imported cases. We model this latent import with an additional Poisson distributed number of cases per observed imported case, with Poisson mean estimated from the data during calibration. 
We call this mean the amplification factor. 

A final aspect, which is not central in this paper but that we mention for completeness, is that we use a geographical SEIR model on county (regional) level, so that the various compartments are geographically defined. Individuals are moved at random between the eleven counties of Norway using a mobility matrix, which is obtained every six hours from the movements of mobile phones, as explained in~\citet{FHImodelPaper2021}. All parameters in the model are however shared between counties. Even if hospitalisation and test data are available at county level, in this paper we use only nationally aggregated data, because of the heterogeneity in the population size among the regions. A very important further aspect is the need to obtain inferential results as rapidly as possible, in at most a few hours, so to be able to publish results quickly just after release of the data update.

\vskip 4mm
\section{Model}\label{sec:model}
\begin{figure}[t]
\centering
\begin{tabular}{c|c}
\begin{tikzcd}[row sep=small]
\arrow[r]&R_{t-1}\arrow[d]\arrow[r]&R_t\arrow[d]\arrow[r]&R_{t+1}\arrow[d]\arrow[r]&\phantom{a}\\
\arrow[r] \arrow[ddr,dashed]\arrow[ddrr,dashed] \arrow[ddrrr,dashed]&\bm s_{t-1}\arrow[r] \arrow[dd] \arrow[ddr,dashed] \arrow[ddrr,dashed]&\bm s_t\arrow[dd]\arrow[r]\arrow[ddr,dashed]&\bm s_{t+1} \arrow[dd]\arrow[r]&\phantom{a}\\
& \phantom{a}&\phantom{a} &\phantom{a}\\
& \bm y_{t-1} &\bm y_t&\bm y_{t+1}&\phantom{a}
\end{tikzcd}&
\begin{tikzcd}[row sep=small]
\arrow[r] &R_{t-1}\arrow[d]\arrow[r]&R_t\arrow[d]\arrow[r]&R_{t+1}\arrow[d]\arrow[r]&\phantom{a}\\
\arrow[r] &\bm s_{t-1}\arrow[r] \arrow[d]&\bm s_t\arrow[d]\arrow[r]&\bm s_{t+1} \arrow[d]\arrow[d]\arrow[r]&\phantom{a}\\
\arrow[r] &\bm z_{t-1}\arrow[r] \arrow[d] &\bm z_t\arrow[d]\arrow[r]&\bm z_{t+1} \arrow[d]\arrow[r]&\phantom{a}\\
& \bm y_{t-1} &\bm y_t&\bm y_{t+1}&\phantom{a}
\end{tikzcd}
\end{tabular}
\caption{\label{fig:graph1}Graphical representation of model~\eqref{eq:ssm1} (left) and its reformulation  (right). The hyper-parameters $\bm\theta$ are not included and can influence all conditional distributions. The dashed arrows illustrate the dependence due to the random delays between infections and testing and possibly hospitalisation.}
\end{figure}
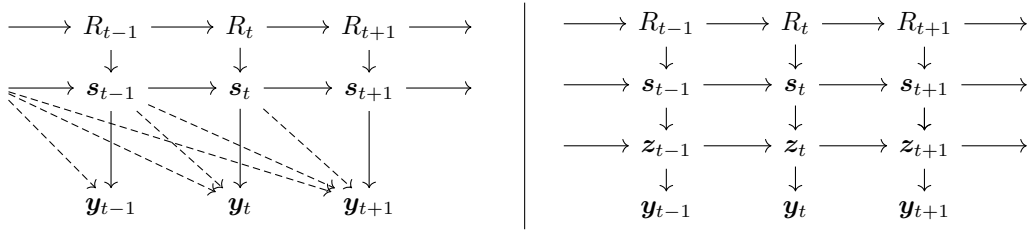

Let $\bm y_t$ be the vector of hospitalisation and test data on day $t$. 
Let $\bm s_t$ be the output vector of compartmental variables in the SEIR model at time $t$, for example the number of individuals in each county who are infected and symptomatic.  
Here we consider the model  generically as an algorithm which outputs the compartmental variables  $\bm s_t$ at each time point $t$. $R_t$ is the unknown reproduction number at time $t$. 
We consider the following state space model:
\begin{subequations}\label{eq:ssm1}
\begin{align}
R_t\sim& P_R(R_{t-1};dR_t);&&\text{for the reproduction number},\\
\bm s_t\sim& P_s(R_t,\bm s_{t-1};d\bm s_t);&&\text{representing the SEIR process},\label{eq:ssm1.s}\\
\bm y_t\sim& H_{y|s}(\bm s_{1:t};d\bm y_t);&&\text{for the hospital and test data}.\label{eq:ssm1.y}
\end{align}
\end{subequations}
To simplify notation, we do not include the dependence of the models on the set of static parameters $\bm\theta$.
The distribution $P_R$ needs to be available analytically and easy to sample from. 
The distribution $P_s$ is assumed to be only available through a computer algorithm and we are only able to simulate from this distribution. In certain situations, this distribution can be available as a huge and complex Markov process. However this is not often the case, for example because of the complexity of the code
or because of the lack of availability of sensitive data, like the mobility matrices in our case. The dimension of $\bm s_t$ is large while $R_t$ is low-dimensional. In this work we consider a common scalar $R_t$ for all counties. 
Note that the data $\bm y_t$ depend on the whole history $\bm s_{1:t}$ making $\bm y_t$ only weakly informative about $(R_t,\bm s_t)$.  This is due to the fact that there is a random delay from transmission to being tested and possibly hospitalised.
A graphical representation of the model is given in the left panel of Figure~\ref{fig:graph1}. 
Our aim is to construct an efficient SMC method for the computation of $p(R_t,\bm s_t|\bm y_{1:T})$ and we are interested in estimating the current status  $(t=T)$, in smoothing ($t<T$) and in  forecasting $t>T$. 

The stochastic process  $\{R_t\}$ is assumed to be Markov. We suggest three  alternative prior models for $P_R$.  Let $\varepsilon_t\sim N(0,\sigma_R^2)$ be independent from all other variables.
The first model is a random walk on log-scale, the second model extends this to an AR(1) structure. In the third model $R_t$ is assumed to be piece-wise constant, with jumps occurring at random.  
\begin{align*}
  R_t=&
 e^{\log(R_{t-1})+\varepsilon_t}\ &&\text{RW, model 1}\\
  R_t=&e^{\mu+a(\log(R_{t-1})-\mu)+\varepsilon_t} &&\text{AR, model 2}\\
 R_t=&
 \begin{cases}
 R_{t-1}&\text{with probability $1-\phi_c$}\\
 R_{t-1}e^{\varepsilon_t}&\text{with probability $\phi_c$.}
 \end{cases}
 &&\text{PC, model 3}
\end{align*}
For all models, we assume a constant start: Until a given time point $D$, $R_t$ is constant and equal to $R_0$. The date $D$ is set to the day when the social-distancing implementation started. Before $D$ the reproduction number was not likely to change significantly. In this paper we set $D$ equal to March 8, 2020, when teleworking started in many companies and universities in Norway. The Norwegian government announced the first package of interventions on March 12 2020. The models above describe the dynamics after $D$. All the static parameters  in the three models must be estimated. 

The proposed method is not dependent on a specific  model~\eqref{eq:ssm1.s}, and would work with any epidemic model. In this work we use the particular SEIR model described in~\citet{engebretsen2019theoretical,engebretsen2020}, and whose algorithm is available though the \texttt{spread} package~\citep{spread2020}.
{\color{myred} 
In particular, this SEIR model has six compartments in each region (county) $i$: susceptible ($S$), exposed and not infectious ($E_1$),
presymptomatic and infectious ($E_2$), infectious symptomatic ($I$), infectious asymptomatic ($I_a $),
and recovered (R). The dynamics is described by the following equations
\begin{equation}\label{eq:SEIR}
\begin{split}
S^i_{t+\delta_t}=&S_t^i-X_{1,t}^i&&X^i_{1,t}\sim\text{Binom}(S^i_t,\beta_t(I^i_t+r_{I_a}I_{a,t}^i+r_{E_2}E_{2,t}^i)\delta_t)\\
E^i_{1,t+\delta_t}=&E_{1,t}^i+X_{1,t}^i-X_{2,t}^i&&X^i_{2,t}\sim\text{Binom}(E_{1,t}^i,\lambda_1\delta_t)
/N_t^i)\\
E^i_{2,t+\delta_t}=&E_{2,t}^i+X_{3,t}^i-X_{4,t}^i&&X_{3,t}^i\sim\text{Binom}(X_{2,t}^i,(1-p_a)\delta_t)\\
&&&X_{4,t}^i\sim\text{Binom}(E_{2,t}^i,\lambda_2\delta_t)\\
I^i_{t+\delta_t}=&I^i_t+X_{4,t}^i-X_{5,t}^i&&X_{5,t}^i\sim\text{Binom}(I^i_t,\gamma\delta_t)\\
I^i_{a,t+\delta_t}=&I^i_{a,t}+X_{2,t}^i-X_{3,t}^i-X_{6,t}^i&&X_{6,t}^i\sim\text{Binom}(I^i_{a,t},\gamma\delta_t),
\end{split}
\end{equation}
where $\delta_t$ is 6 hours. 
We assume random mixing in each county in each 6 hour period, and move individuals between counties at the end of each such period according to a mobility matrix. In ~\citet{FHImodelPaper2021} mobile phone data are used to estimate such mobility matrices. These matrices report the number of individuals moving from county A to county B during each period, which we pick at random among the currently present in A, but favour the residents of B to return to B, to capture commuting. This rules makes the computational complexity of the SEIR model quadratic in the number of counties, due to the need for storage of both current visited location and residence of individuals. The reproduction number $R_t$ is related to $\beta_t$ through the equation
\[
R_t=\beta_t\left(\frac{1-p_a+p_ar_{I_a}}{\gamma}+\frac{(1-p_a)r_{E_2}}{\lambda_2}\right).
\]
}

Finally we describe the likelihood model for the data~\eqref{eq:ssm1.y}.
A main difficulty is the link between the latent process $\{\bm s_t\}$ and the observation process $\{\bm y_t\}$ because of the unknown stochastic delays between infection and observation time, making the computation of $H_{y|s}(\bm s_{1:t};d\bm y_t)$ hard. We introduce an auxiliary process  $\bm z_t=(\bm z^H_{t,0:L^H},\bm z^T_{t,0:L^T})$ with two components, one dedicated to the hospitalisation and the other to the test data. The auxiliary variable $z_{t,v}^H$ is defined as the number of individuals who are infected at time $t$ and hospitalised $v$ days later. The time lag $v$ is assumed to vary in $\{0,...,L^H\}$, for some appropriate $L^H$.
Similarly $z_{t,v}^T$ is the number of individuals who are infected at time $t$ and tested positive $v$ days later.
We rewrite~\eqref{eq:ssm1.y} as
\begin{subequations}\label{eq:ssm2.y}
 \begin{align}
\bm z_t\sim& G(\bm z_{t-1},\bm s_t;d\bm z_t),\\
\bm y_t\sim& H_{y|z}(\bm z_{1:t};d\bm y_t),
\end{align}
\end{subequations}
where $G(\bm z_{t-1},\bm s_t;d\bm z_t)$  is a Markov transition distribution assumed to be easy to simulate from. Now $H_{y|z}(\bm z_{1:t};d\bm y_t)$ is easy to compute. 
In more details,  define $y^H_t$ to be the number of daily Covid-19 admissions to hospital. An individual who is infected at time $t$ is hospitalised with probability $p_u^H$ at time $t+u$. The time lag $u$ from infection until hospitalisation is assumed to follow a
 discrete distribution on the integers $\{0,1,...,L^H\}$. Let $\rho_u^H$ be the probability of delay $u$.
To  structure the model further, we attach to each infected individual its potential time lag until hospitalisation.
So $z_{t,u}^H$ is the number of individuals infected at time $t$ and who could possibly be hospitalised  $u$ time-units later. Let $I_t$ be the total number of infected individuals at time $t$, which is available through the SEIR model as a component of $\bm s_t$. 
We can formulate the model as
\begin{equation}\label{eq:obs}
\begin{split}
z_{t,0}^H|I_t\sim&\text{Binomial}(I_t,\rho^H_0);\\
z_{t,u}^H|I_t,z^H_{t,0:u-1} \sim&\text{Binomial}\left(I_t-\sum_{v=0}^{u-1}z_{t,v}^H,\frac{\rho^H_u}{1-\sum_{v=0}^{u-1}\rho^H_v}\right)&&u=1,...,L^H;\\
y_t^H\sim&\text{Beta-Binomial}(\sum_{u=0}^Lz^H_{t-u,u},\alpha^H,\beta_t^H),
\end{split}
\end{equation}
{\color{myred}where we here consider a Beta-Binomial distribution for a patient being hospitalised to take into account variability in hospitalisation between regions. 
The $\beta_t^H$ parameter is specified indirectly through a time-varying probability $p_t^H$ such that $\beta_t^H=\alpha(1-p_t^H)/p_t^H$ where $p_t^H$ is predefined using the age-structure of the individuals having tested positive. By storing and sequentially updating the quantities $\sum_{u=0}^Lz^H_{t-u,u}$ as well, we obtain a first order Markovian state space structure as illustrated in the right panel of Figure~\ref{fig:graph1}.}

Regarding the test data, as in \citet{FHImodelPaper2021} we consider the probability that an infected case is tested by means of a PCR test. We ignore that tests can lead to false positive responses. The logit of this probability is assumed to be linear in the total number of daily tests $N^T_t$ in day $t$ in addition to a time independent intercept. We write the detection probability $\rho^T_t$ at time $t$  as
\begin{align}
\rho^T_t=\frac{\exp(\pi_0+\pi_1 N^T_t)}{1+\exp(\pi_0+\pi_1 N^T_t)}.\label{eq:pi_test}
\end{align}
The time lag between infection and testing is assumed to follow a
 discrete distribution on $\{0,1,...,L^T\}$ for an appropriate $L^T$.  The approach for handling this delay is exactly as for the hospital incidence, with $p^T_t$ now playing the role of $p_t^H$.
We introduce a new set of auxiliary variables for the test data $\{z_{t,u}^T\}$, similarly to the ones introduced for the hospitalisation data. Defining $\bm z_t=(\bm z^H_{t,0:L^H},\bm z^T_{t,0:L^T})$, we are within the model formulation~\eqref{eq:ssm2.y}.

\vskip 4mm
\section{Sequential Monte Carlo}\label{sec:smc}

Let $\bm x_t=(R_t,\bm s_t,\bm z_t)$.
Our aim is to perform inference on the whole set of latent variables $\bm x_{1:t}=(\bm x_1,...,\bm x_t)$ as well as on static hyper-parameters $\bm\theta$ at each time-point $t$, by means of the posterior distribution $p(\bm x_{1:t},\bm\theta|\bm y_{1:t})$. 
A description of the SMC algorithm with resampling at each step is  given in Algorithm~\ref{alg:smc}. See also \citet{chopin2020introduction}, which contains more general algorithms. Although in this paper we focus on inference for $R_t$, also $\bm s_t$ will be of interest.
In our setting, the main computational burden is the sampling from $P_s(R_t,\bm s_{t-1};d\bm s_t)$ which has been parallelised in our implementation. {\color{myred}For resampling, residual resampling~\citep{liu1998sequential} has been applied. However, the resampling step is both hard to parallelise and requires message passing, resulting in that a too high number of cores can decrease performance.}

\begin{algorithm}[t]
\caption{Auxiliary SMC with resampling at each time step. Operations involving index $b$ must be performed for $b=1,...,B$. Here $P_t^x$ denotes the transition distribution for $\bm x_t$, while $Q_t$ is the proposal distribution for $\bm x_t$. The indices $A_t^{1:B}$ defines the ancestral particles at time $t$ after resampling.}
\label{alg:smc}
\begin{algorithmic}[1]
\State $\bm x^b_0\sim Q_0(d\bm x_0)$\Comment{Proposal at time 0}
\State $w^b_0=\frac{P_0(d\bm x^b_0)H(\bm y_t|\bm x^b_0)}{Q_0(d\bm x^b_0)}$,\quad $W^b_0=w^b_0/\sum_{m=1}^Bw^{m}_0$\Comment{Calculating weights}
\State $\ell_0^B=\frac{1}{B}\sum_{b=1}^Bw^b_0$\Comment{Estimate of $p(\bm y_0)$}
\For{$t=1$ to $T$}
        \State $A_t^{1:B} = \text{resample}(W^{1:B}_{t-1})$\Comment{Resampling}
\State $\bm x^b_t\sim Q_t(\bm x^{A_t^b}_{t-1},d\bm x_t)$\Comment{Proposal at time $t$}
\State $w^b_t=w^b_{t-1}\frac{P_t(\bm x^{A_t^b}_{t-1},d\bm x^b_t)H(\bm y_t|\bm x^b_t)}{Q_t(\bm x^{A_t^b}_{t-1},d\bm x^b_t)}$, \quad $W^b_t=w^b_t/\sum_{m=1}^Bw^m_t$\Comment{Calculating weights}
\State $\ell^B_t=\frac{1}{B}\sum_{b=1}^Bw^b_t$\Comment{Estimate of $p(\bm y_t|\bm y_{t-1})$}
\EndFor
\end{algorithmic}
\end{algorithm}

The SMC algorithm is by design sequential so that by storing values of $\bm x_t$ obtained at the previous day, updates can easily be performed as new data arrive.  
A main challenge here is that the state $\bm x_t$ at time $t$ heavily depends on \emph{future} observations $\bm y_{t+h}$ because of the delay in hospitalisation. Although the reformulated model reduces immediate dependence, there are still strong correlations backwards, as illustrated in Figure~\ref{fig:graph1}. There are clearly possibilities to develop more efficient proposal distributions, despite the availability of the SEIR model only as a computer algorithm. Because the main purpose of this paper is to investigate the  utility of SMC methods for the estimation of daily varying reproduction numbers, we use only simple bootstrapping proposals. We do however take into account the delay aspect by
using fixed lag smoothing, using data ahead of current time, that is
\[
\widehat R_{t|t+l_t}=E[R_t|\bm y_{1:t+l_t}]\approx\frac{1}{B}\sum_{b=1}^BR_t^b
\]
where $\{R_t^b\}$ are simulations of $R_t$ based on data up to time point $t+l_t$.
Fixed lag smoothing with a lag of $l_t=24$ days has been used in the Covid-19 runs in this paper.
At the end of the time-series, $l_t=\min\{T-t,l\}$ is used. The estimates on the last days will be more uncertain.

\vskip 4mm
{\color{myred}
\subsection{Parameter estimation}
}
Algorithm~\ref{alg:smc} assumes that the parameters $\bm\theta$ are known. 
Now we describe Bayesian inference for some of the static parameters. 
We denote by $\bm\theta_R$ the set of parameters in the model for $\{R_t\}$, $\bm\theta_s$ the ones in the $\{\bm s_t\}$ process, and $\bm\theta_y$  the parameters appearing in the data model. 
As mentioned, some of these parameters are fixed based on other data sources, and here for simplicity we do not propagate their uncertainty. For other parameters, sequential updates of their estimates are desirable.
In principle, all parameters $\bm\theta$ could be included as part of the state vector $\bm x_t$ where the propagation of these static components just keeps them fixed. However, repeated resampling will quickly give degenerate samples for these parameters. 

{\color{myred}
A review of parameter estimation in SMC is given in~\citet{kantas2015particle}. Off-line methods such as Particle MCMC~\cite[PMCMC,][]{andrieu2010particle} have proven to be very effective in many applications, but require repeated runs of the SMC routine. Although much smaller number of particles can be applied in such settings, some experiments with our models indicate that at least 250 particles are necessary, in which case one run uses about 10 minutes using 4 cores on Linux server and more cores did not help much in this case. Some experiments with an implementation of the Particle Metropolis-Hastings algorithm, based on the pseudo-marginal method by~\citet{andrieu2009pseudo}, is reported in supplementary section~\ref{sec:Spar}. We will however focus on online methods for parameter estimation here.}

In cases where sufficient statistics $\bm v_t(\bm x_{1:t})$ for the parameters are available, the SMC algorithm can be easily updated to target $p(\bm x_{1:t},\bm v_t(\bm x_{1:t})|\bm y_{1:t})$ instead~\citep{fearnhead2002markov,storvik2002particle}. This is the case for the $\bm\theta_R$ parameters. Simulations of $\bm\theta_R$ at each time point can then be obtained from $p(\bm\theta_R|\bm v_t(\bm x_{1:t}))$ which then again can be used to obtain new samples of $\bm x_{t+1}$ (step 6 in Algorithm~\ref{alg:smc}). A crucial step is that $\bm v_t(\bm x_{1:t})$  can be recursively updated. 
Section~\ref{sec:par.R} in the supplementary material gives details on how $\bm\theta_R$ can be updated by this approach.  
Note that these methods can suffer from the degeneracy problem \citep{andrieu2005line}. In supplementary section~\ref{sec:Spar} we validate the parameter estimates obtained by this procedure both through comparisons between different runs and by using the samples obtained by the Particle Metropolis-Hastings algorithm.


\vskip 4mm
\section{Results}\label{sec:res}

\subsection{Norwegian Covid-19 data}
\begin{figure}
    \centering
    \includegraphics[width=\textwidth]{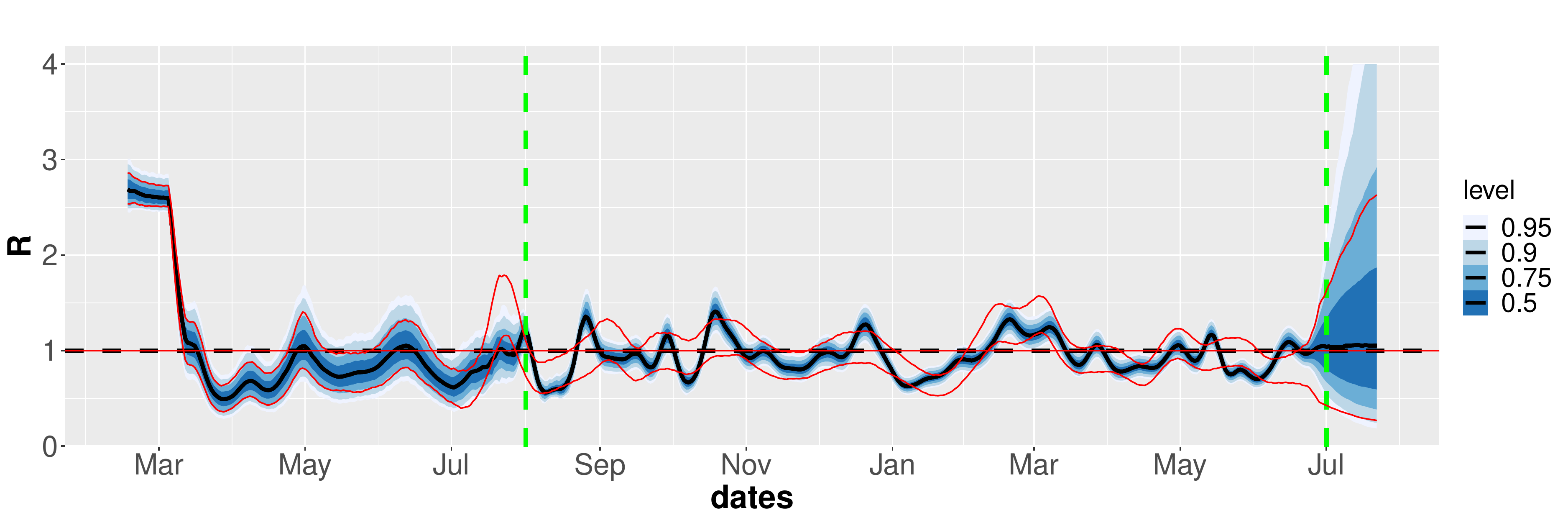}\\
    \includegraphics[width=\textwidth]{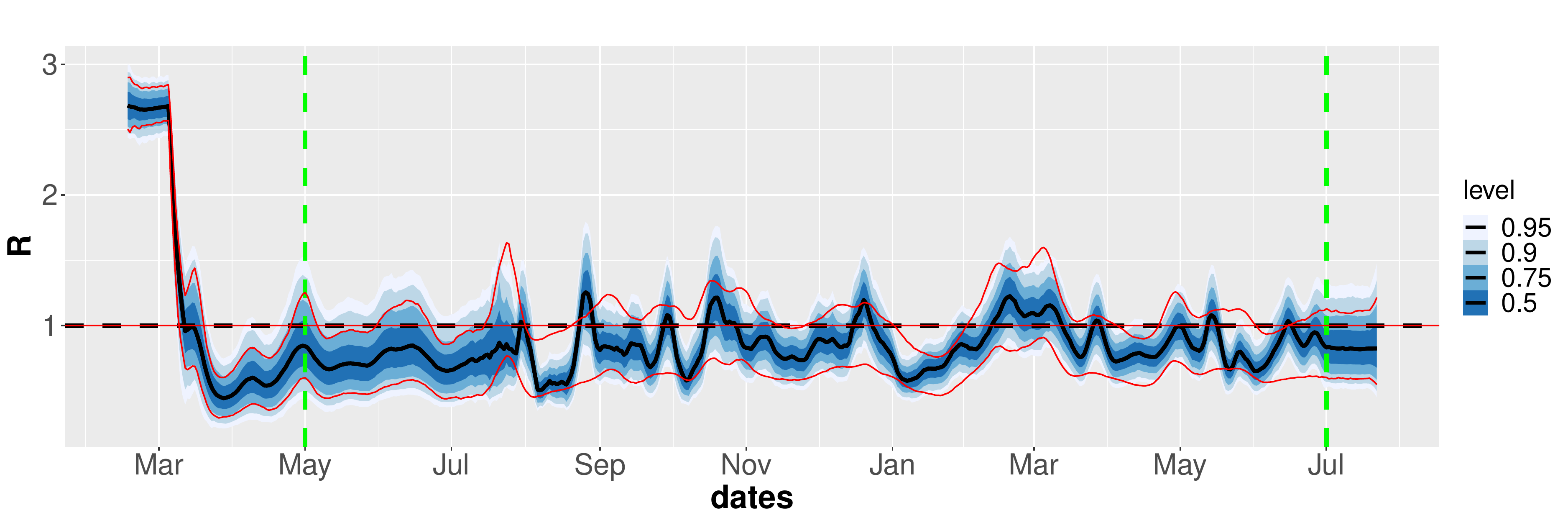}
    \includegraphics[width=\textwidth]{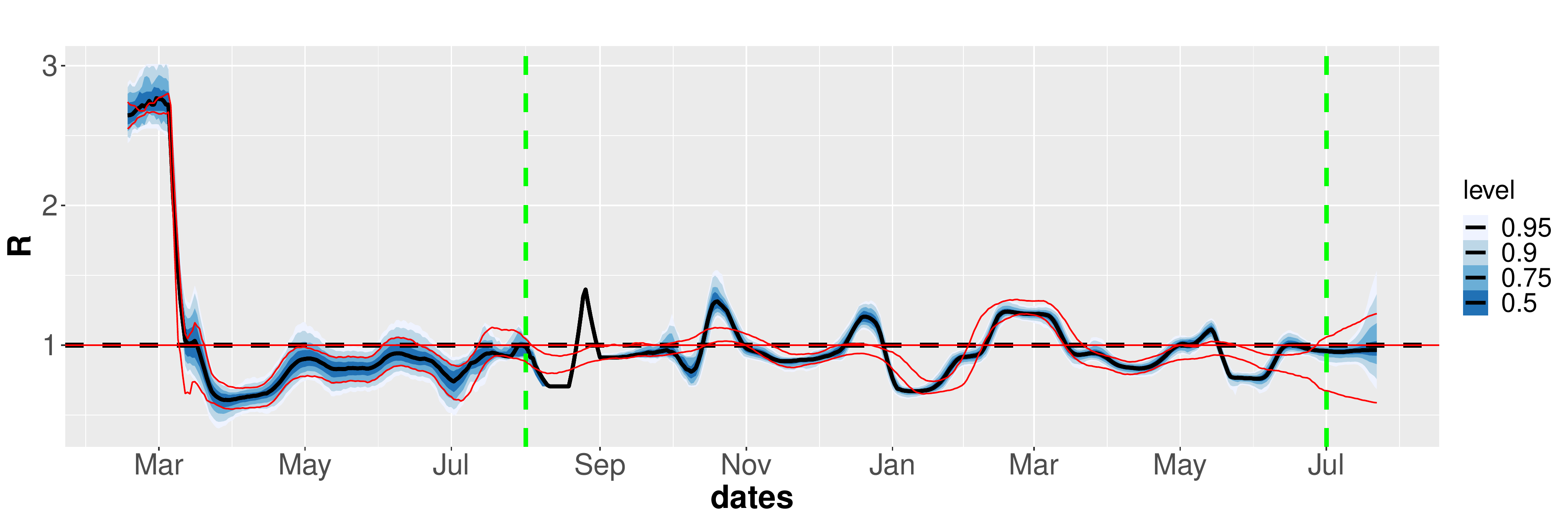}
    \caption{\label{fig:R_t_week}Estimates of the weekly averaged instantaneous reproduction number 
    $\frac{1}{7} \sum_{u=0}^6R_{t-u}$ based on the random walk (RW, upper), autoregressive (AR, middle) and piece-wise constant model (PC, lower). We used hospital data from March 8 2020 to July 1 2021, test data from August 1 2020 to July 1 2021. We used $B=20\, 000$ particles. 
    Corresponding not-averaged daily estimates are given in Figure~\ref{fig:R_t_daily}. The first vertical green line corresponds to the date at which test data is included in the analysis, the second green vertical line corresponds to the last date with observations, after that Bayesian prediction is performed. Blue shadow bands indicate posterior uncertainty quantiles. 
    The red curves correspond to 50\% median-centred credibility intervals only based on hospitalisation data.}
\end{figure}

{\color{myred} For the analysis of the Norwegian Covid-19 case, we have used the 11 counties as our spatial scale. Mobility data and imported cases are used at regional scale while hospital incidence data and test data are used at national scale. Also the $\{R_t\}$ prior process is assumed to be common for all regions.}
The hospital incidence data are from the Norwegian national Beredt-C19 registry and the Norwegian Intensive Care and Pandemic Registry, and the test data are from the Norwegian Surveillance System for Communicable Diseases  (MSIS registry). The reproduction number is assumed constant until March 7 2020. 
The results are based on data up to July 1, 2021 with test data included from August 1, 2020, when testing capacity in Norway was scaled as needed, and after which testing criteria had become rather stable. The number of parameters is large: the dimension of $\bm s_t$ is 3157. In addition we have the reproduction numbers $R_t$ and the auxiliary variables $\bm z_t$'s. The prior distributions assumed for the parameters involved in the dynamics of $R_t$ are given in the first column of Table~\ref{tab:par.est}.
\begin{table}[t]
    \centering
    \begin{tabular}{l|ccc}
    &\multicolumn{3}{c}{Model}\\
   Prior&RW&AR&CP\\
\hline
$a\sim\text{N}(0.5,0.25)$&&0.563 (0.430,0.659)  & \\
$\frac{1}{\sigma^2}\sim\text{Gamma}(2.4,0.28)$& 0.176 (0.157,0.202) &0.340 (0.300,0.477)&0.574 (0.485,0.678)\\
$\phi\sim\text{Beta}(1,9)$&&&0.158 (0.119, 0.198) \\
\hline
MLIK&-3028.20&-3008.68& -3020.69\\   
MLIK-no test&-1252.15&-1221.09& -1230.71 \\
\hline
    \end{tabular}
    \caption{\label{tab:par.est}Posterior medians and 95\% credibility intervals (in parentheses) for hyper-parameters of the three models of the  $R_t$ dynamics. Although the prior for $\sigma$ is defined through the precision, the numbers are for $\sigma$ itself. The two last rows of the table give the value of the marginal log-likelihoods based on both data sources (MLIK) and on hospitalisation data only (MLIK-no test).}
    
\end{table}
{\color {red}When running the model, we did not estimate the parameters in the SEIR model~\eqref{eq:SEIR} nor the parameters $\pi_0$ and $\pi_1$ in~\eqref{eq:pi_test}, because they were estimated separately as described in~\citet{FHImodelPaper2021}. Also the parameters related to observations in equation~\eqref{eq:obs} were pre-estimated through other data sources.}
All further details on the model are given in supplementary material section~\ref{sup:parset}. Each run is based on $B=20\,000$ particles. One run of the 500 days considered here, using a linux server with 128 cores, took approximately 5 hours, which is appropriate for practical real-time purposes. 
 Figure~\ref{fig:R_t_week} shows our estimates of $R_t$, with a 7-precedent-days moving-average smoothing (daily estimates are given in the supplementary material Figure S3), using the three considered models for the reproduction number. 
 Posterior medians and symmetric 95\% credibility intervals of the estimated parameters are reported in the second column of Table~\ref{tab:par.est}.
 {\color{myred} The autoregressive model for $R_t$ was simplified by fixing $\mu=0$: when we estimated $\mu$, we obtained an estimate very close to zero, but also some difficulties in the estimation of $\sigma^2$. In addition, the  marginal log-likelihood (MLIK) was slightly higher. We therefore opted for the simpler autoregressive model.}
 
In Figure~\ref{fig:R_t_week} we see that the different models lead to rather similar estimates for the time dynamics of $\{R_t\}$. The autoregressive model (central panel) seems to have more uncertainty. On the other hand, this model is to be the preferred in terms of  MLIK (Table~\ref{tab:par.est}, last rows). Note that all models capture quickly the dramatic reduction in transmission in mid March 2020 when Norway had the first lockdown, which was implemented between March 9 and March 14 2020. In the second half of April 2020, society was re-opened, including schools and kindergarten, the 2-meters distance rule was reduced to 1-meter: $\{R_t\}$ appears to increase to around 1. A new peak appears in the end of July and beginning of August 2020, in correspondence with the end of the Norwegian vacations, the returning of Norwegians from abroad and the arrival in Norway of tourists. In particular there have been several isolated clusters, for example in two cruise ships and the return of students to university campuses. These clusters were well controlled by contract tracing, and the reproduction number dropped rapidly below 1. The autumn 2020 was also characterised by a series of larger local outbreaks, which were rapidly controlled. The growth of $R_t$ starting in October 2020 was also due to local outbreaks, and the incoming winter-season which affects the viral transmission. The number of cases was so large that contact tracing became less effective. The Norwegian government imposed a second national intervention, first in the end of October 2020 and then again in early November 2020. {\color{myred}This reduced the reproduction number again below 1, in two pushes, where we can see that the second strengthening of the intervention was in fact needed for this purpose.  We can see that the autumn 2020 interventions allowed the reproduction number to fall from approximately 1.5 to below 1 in about three weeks. Interestingly, we see a peak of $R_t$ just around Christmas 2020, in connection with vacation travel and intensive shopping. January 2021 marks the arrival of the alpha variant of the virus, which was more transmissible and which increased the risk of hospitalisation. The alpha variant was predominant in Norway at the end of March 2021, and we see $R_t$ just below 1.5 again. This increase happened despite at the end of January Norway introduced the strictest lockdown rules during the whole pandemic so far, including essential closure of all borders, and vaccination of the elderly started. During March and April 2021, $R_t$ started to decrease again, to remain below or around 1 until the middle of May 2021. Governmental interventions were reduced from May 2021, and the reproduction number stabilised around 1. At the end June 2021, approximately 50\% of the adult Norwegian population had been vaccinated at least once, and approximately 30\% twice. The effective reproduction number $R_t$ reflects the effect of vaccination which is included in the SEIR model.}
It is remarkable how the estimated reproduction number quantify the history of the epidemic so precisely.
Another feature is the cyclic behaviour of $R_t$, with a drop following an increase. The AR prior model on log scale also attracts towards 1. In Norway this is expected because of the rapid intervention strategy of the government (named "control") whenever $R_t$ was growing rapidly above 1, and the rapid reopening when the epidemic appeared under control. Local outbreaks were frequent, also visible in the raw data, but they were rapidly controlled by appropriate contact tracing and other successful local interventions.

The red lines in Figure~\ref{fig:R_t_week} give 50\% centred credibility intervals based only on hospitalisation data. We investigate in this way the value of the test data as a second source of information. Because we used test data only from August 1 2020, the estimates are essentially identical until then. 
After August 1 2020, the estimates based only on hospital data  are smoother, indicating that the test data contain information about transmission (and its change) that is not transferred to the hospitalisation. One reason for this is that the younger generations have been infected in the autumn more than the elderly ones, who are most at risk for hospitalisation. The test data also contribute to a more precise estimate of the daily prevalence of infected in Norway. We also observe some misalignment in time between the two estimates, probably because the time lags are not stationary, while we assume them constant during the whole epidemic.  

\begin{figure}
    \centering
\includegraphics[width=\textwidth]{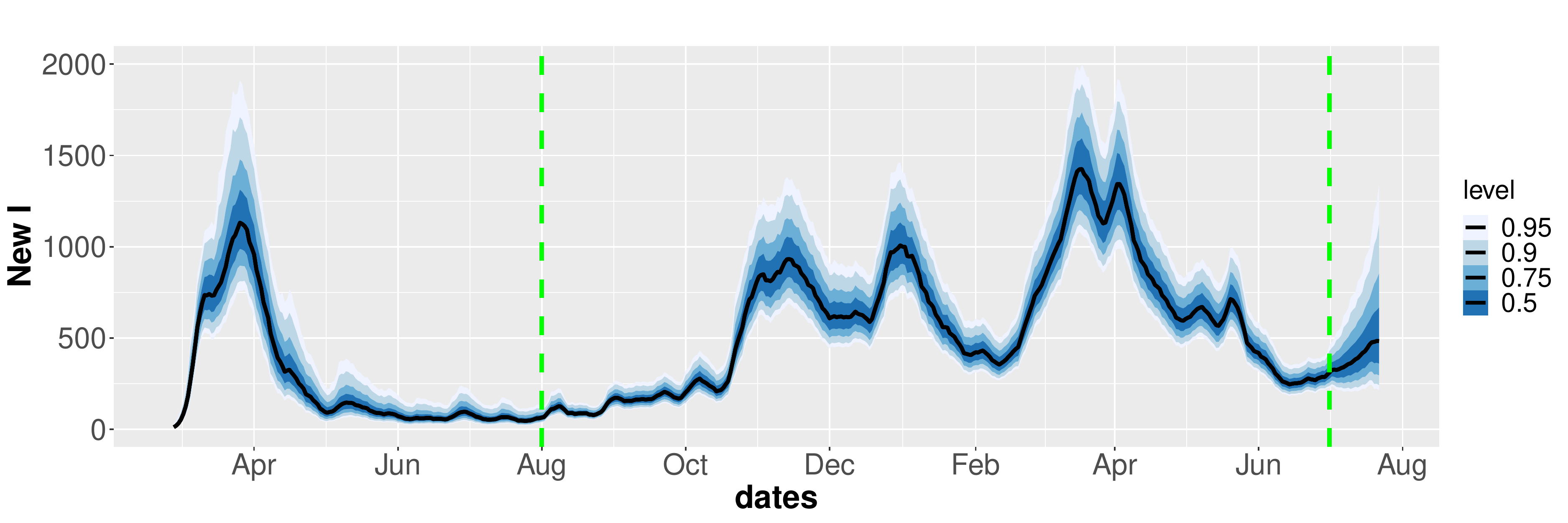}
    \caption{\label{fig:pred.I}Predicted number of newly infected based on the autoregressive model in Figure~\ref{fig:R_t_week}. The estimates after the last dashed green line are predictions 3 weeks ahead.}
\end{figure}

The last green horizontal line in Figure~\ref{fig:R_t_week} corresponds to the last day with data, July 1, 2021, after which we perform a three weeks prediction, by simply running the model forward in time. The predictions for the three models have similar means but differ in uncertainty. Both the random walk and the piece-wise constant model are non-stationary, which explains the high increase in uncertainty after the last observation point. On the other hand, the autoregressive model shows a more stable prediction performance, as expected. {\color{myred}Figure~\ref{fig:pred.I} gives predicted number of new daily infected cases. Before the last green dashed line these credibility intervals give predictions based on the observed data, after this line, the predictions are obtained by running the SEIR model forward three weeks in time using the predicted reproduction numbers. When predicting forward in time, mobility matrices, imported cases and total number of tests are needed as input. We here re-use the values in the previous 21 days in the forecasts. We see the three waves which hit Norway in March/April 2020, from November 2020 to January 2021 and in March 2021. The number of cases is estimated around 1000 per day in the peaks (1300 in the third wave). The number of cases would in fact grow during July 2021, as here correctly predicted. 
} 


Table~\ref{tab:par.est} provides the estimates of the parameters in the three dynamic models of $\{R_t\}$. How these estimates are learned over time is shown in Figures~\ref{fig:R_rw_par}-\ref{fig:R_cp_par} in the supplementary material. {\color {red} The plots also allow a comparison of the prior (which is the first time point to the left) and the final posterior estimate (last time point on the right). The estimates stabilise nicely. We report about some limited validation of the parameter estimates in section~\ref{sec:Spar}.} Note that the variance of the random walk dynamics is estimated to be smaller than the one of the autoregressive model, as is also clear in Figure~\ref{fig:R_t_week}. On the other hand, the variance related to the piece-wise constant model is considerably smaller even if the variability seems to be smaller in Figure~\ref{fig:R_t_week}. This is due the fact that for most time points there are no discontinuities while when changes occur there may be large discontinuities. 

\begin{figure}
\centering
\begin{tabular}{cc}
\includegraphics[width=\textwidth]{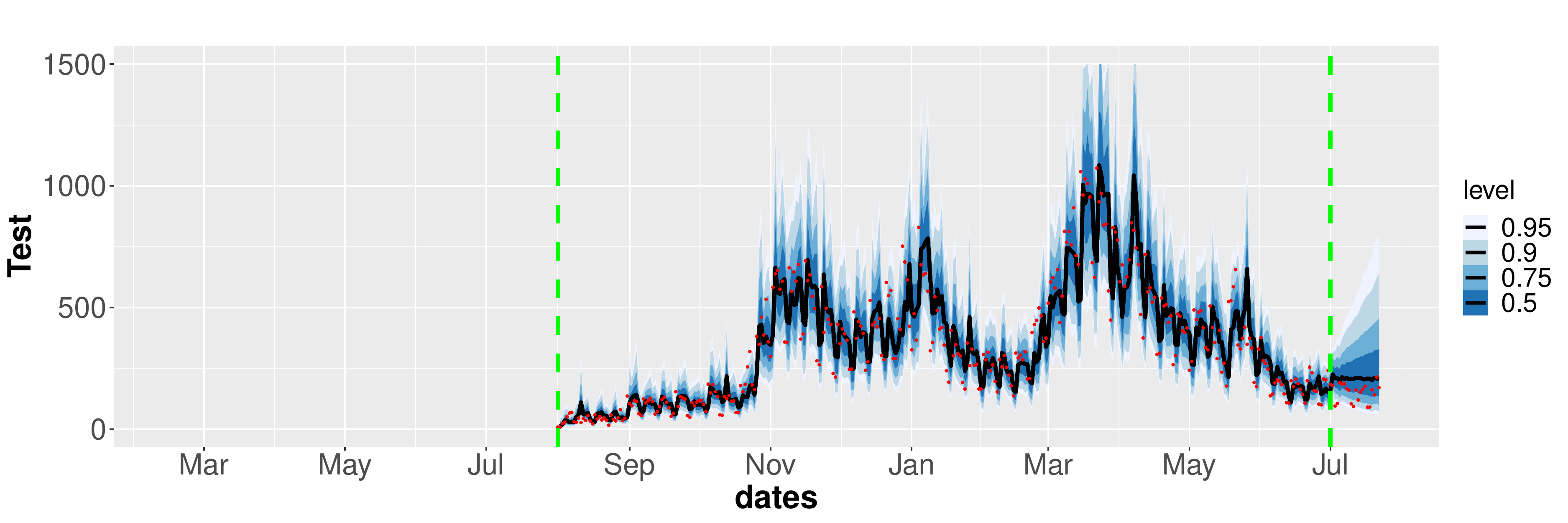}\\
\hspace*{0.2cm}\includegraphics[width=0.985\textwidth]{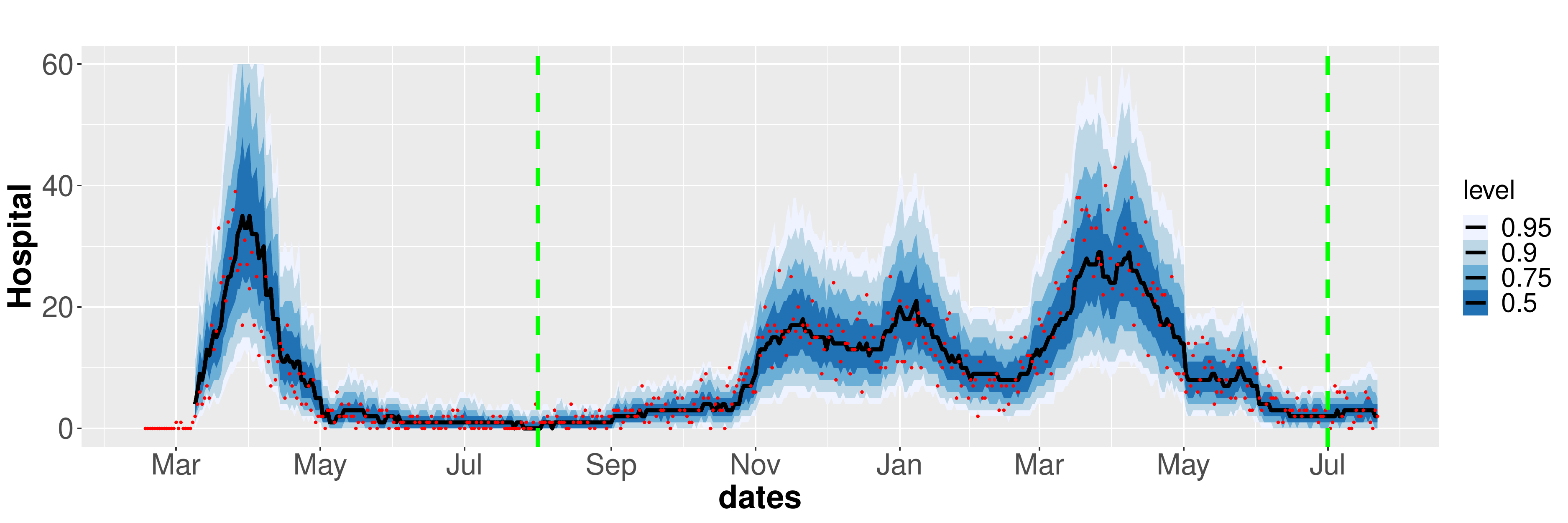}
\end{tabular}
\caption{\label{fig:fit_data}Predicted (blue shadows representing the quantiles of the posterior predictive distribution) and observed daily hospitalisation incidence  and laboratory-confirmed positive tests incidence (red dots) based on the autoregressive model. }
\end{figure}

In Figure~\ref{fig:fit_data} we use the estimated parameters, including the instantaneous reproduction numbers, to simulate the daily hospitalisation incidence and the daily number of positive tests. We propagate uncertainty and produce probabilistic estimates, which we compare with the actual data. These plots show that we are able to fit both data sources well. 
We note the weekly structure in the test data. {\color{myred}These plots also show three weeks ahead forecasts. The superimposition of the actual data of these three weeks, which were not used in the analysis, show that predictions were good.}   


\hskip 4 mm
\subsection{Sensitivity analysis}
There are several parameters that are fixed in the current implementation:
\begin{itemize}
\item Hyper-parameters in the prior model for  $\{R_t\}$, see Table~\ref{tab:par.est}.
\item Parameters related to the observations:  the age-dependent probabilities of being hospitalised and detected positive by testing, and the parameters describing the distribution for delay from infection to testing or hospitalisation.
\item Several parameters inside the SEIR model.
\item Number of particles used in the SMC runs.  
\end{itemize}
We did not perform a systematic sensitivity analysis, but focus on the SMC design parameters. 
Figure~\ref{fig:comp_prior} shows posterior median estimates of the instantaneous reproduction numbers for several different prior settings for the autoregressive model. The figure indicates that the estimates of $\{R_t\}$ are not sensitive to the tested prior settings (credibility intervals also show similar results).

Next we studied the importance of the number of particles. Figure~\ref{fig:R_t_week.B} in the supplementary file compares results for the autoregressive model based on $B=20\,000$ vs. $B=2\,000$ particles. The difference in the estimated marginal likelihoods is small compared to the differences between models (about 5.0). This figure (and other similar ones, not included) shows that the results are quite stable and that $B=2\,000$ might have been sufficient for estimation of $\{R_t\}$. However, if interest is also in the latent structure $\bm s_t$ or marginal likelihood values, more particles have shown to be necessary.

\hskip 4 mm

\begin{figure}
    \centering
    \includegraphics[width=\textwidth]{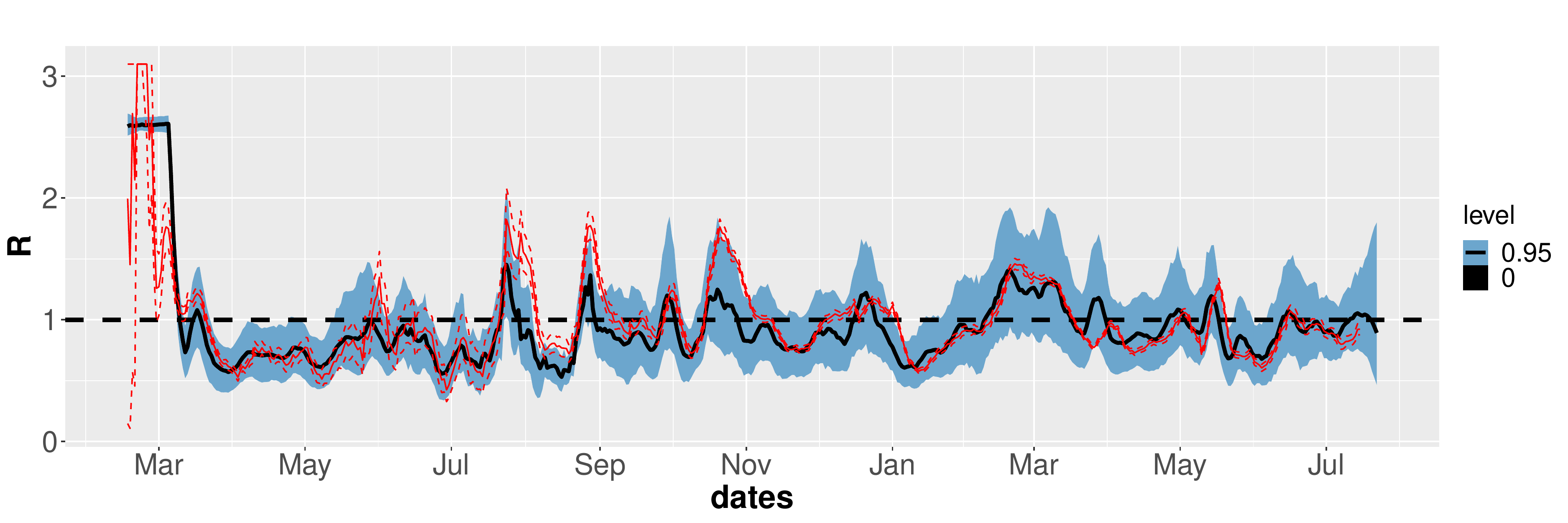}
    \caption{\label{fig:EpiEstim} Estimates of $R_t$ using the autoregressive SMC model and the SEIR based only on test data (blue band) and estimates obtained from \texttt{EpiEstim} (red lines).}
\end{figure}

{\color{myred}
\subsection{Comparison with EpiEstim}
EpiEstim~\citep{cori2013new} is a popular method for estimation of the reproduction number $R_t$ based on case incidence as well as imported cases. It does however not allow for the incorporation of multiple data, so that hospital data are not taken into account, nor mobility. In Figure~\ref{fig:EpiEstim} we compare estimates of $R_t$ obtained by our SMC approach, though only using test data and this time from the start of the epidemic, with the results obtained by using the \texttt{EpiEstim} package in R~\citep[][version 2.2.4]{Cori2021}. For \texttt{EpiEstim} we assumed a serial interval (the time between the onset of symptoms of a primary case and the onset of symptoms of secondary cases) with mean 7.5 days and a standard deviation of 3 days.  We see a very good agreement between the two estimated curves in general. However, the confidence bands are much narrower for  \texttt{EpiEstim}. There are some differences between the estimates in March 2020 and during the summer 2020 and in November 2020. }

\hskip 4 mm

\begin{figure}
    \centering
    \includegraphics[width=\textwidth]{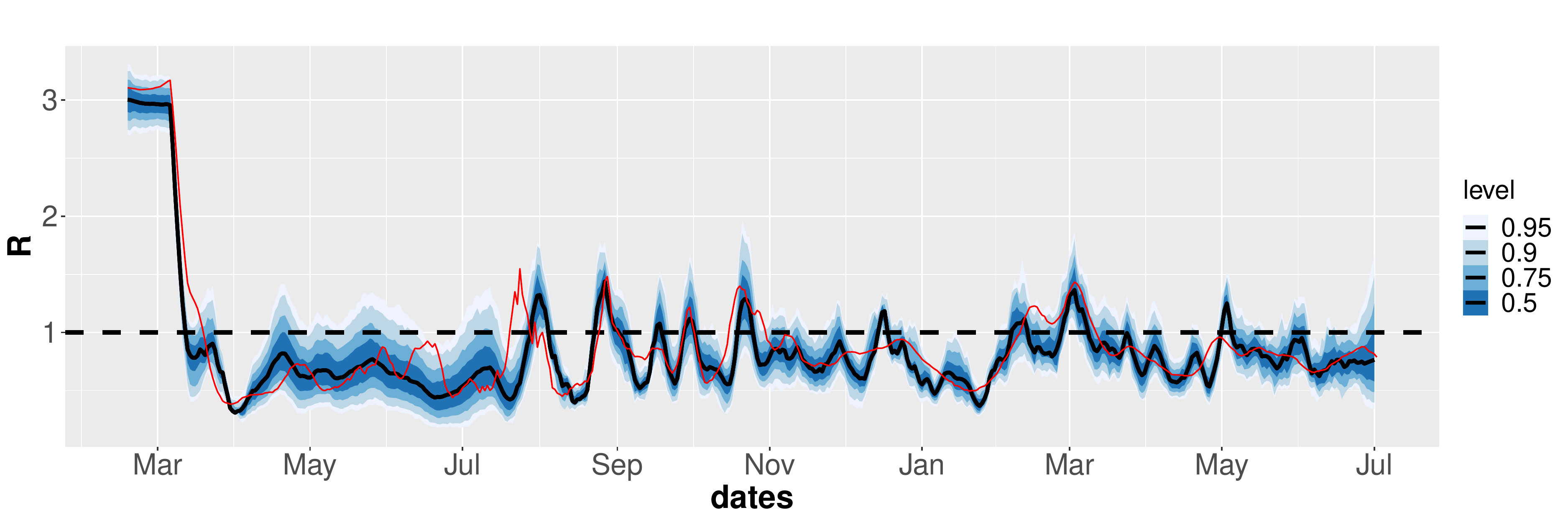}
    \caption{\label{fig:sim} Simulation experiment. Estimates of  $\frac{1}{7} \sum_{u=0}^6R_{t-u}$  using $B=20\,000$ (blue) based on one simulated data set. Same setup as Figure~\ref{fig:R_t_week},   hospitalisation data from March 12 2020 to July 1 2021, test data from August 1 2020 to July 1 2021. The red curve corresponds to the assumed true $\{R_t^*\}$ process.}
\end{figure}

{\color{myred}
\subsection{Simulated data}\label{sec:sim}
Based on an estimate $\{R^*_t\}$ obtained by the AR model from the real data, we simulated $(\bm s_t,\bm z_t,\bm y_t)$  from the model for all $t$. Twenty independent data sets were generated and for each case the simulated set $\{y_t\}$ was used for estimating $\{R_t\}$. Figure~\ref{fig:sim} shows the results from one of these simulated data sets, demonstrating that the method is able to capture the main structure quite well, although in periods where the true $R_t^*$ is considerably unstable (July/August 2020) estimates are worse. 
Figure~\ref{fig:R_sim_rep} in the supplementary material shows results based on all 20 data sets, demonstrating the strong information content in such type of data about the $\{R_t\}$ process.}

\hskip 4 mm

\section{Discussion}\label{sec:disc}

 A time-varying transmissibility allows to quantify the effects of interventions and changes in the behaviour of people, in real-time. This is of key importance to policy makers, as the interventions often have immense societal and health costs. Understanding, while an epidemic develops, whether the implemented interventions are sufficient or not, or if interventions could be lifted, is essential. The Norwegian government's strategy was to control the epidemic, and this was achieved by multiple national and local non-pharmaceutical interventions, which are reflected in the temporal variations of the reproduction number. Our approach is the first which allows to monitor a daily-varying reproduction number when using a complex compartmental model informed by multiple streams of data. 
 The fact that our estimates of $R_t$ react rapidly to changes in the test data, 
 {\color {red}
 means that the situation is captured only with a delay given by the generation time of the disease under study and the time gap between transmission and testing. For Covid-19, this amounts to about a week, because of a generation time of about 5 days and a delay between transmission and test of about 2 days.}
Picking up an exponential growth ($R_t>1$) before the epidemic grows out of control is essential for surveillance. The possibility of our method to validate the efficacy of contact tracing, to lead back $R_t$ to below 1, or not, is also very important. 

We have shown how daily reproduction numbers and the latent compartment-wise populations in an SEIR model can be put into a state space model, so that an SMC technique for inference can be used. Obtaining unbiased estimates of the marginal likelihoods also makes it possible to do parameter estimation within a particle MCMC framework, although more work is needed here to make this computationally efficient. Compared to a parallel effort using Approximate Bayesian Computation (ABC)~\citep{FHImodelPaper2021}, the SMC approach is much faster and also easier to modify with respect to model changes, confirming the findings in \citet{birrell2020efficient}. Our implementation is modular, does not depend on the specific epidemic model (here SEIR), so that alternatives can easily be tested.

So far only simple bootstrap filters have been applied. This can be improved by utilising more efficient proposals, alternative algorithms such as the resample-move algorithm~\citep{gilks2001following} which was used in~\citet{birrell2020efficient} or the recent promising ideas of iterated auxiliary filters and twisted models~\citep{guarniero2017iterated,heng2020controlled}. We expect these approaches to be very useful when we expand the SEIR model to have different reproductions numbers in each of the eleven Norwegian counties. 

We compared three simple dynamics for $R_t$ and found that the autoregressive model was slightly better than the others. More work needs to be done here to compare the models in terms of prediction. In our approach it is easy to predict the future hospitalisation incidence and the number of positive cases tested. {\color{myred}Note however that such simple dynamic models for $\{R_t\}$ are mostly suitable for now-casting and for short term forecasting, because of the lack of stationary that interventions and feedbacks imply.}
There are then interesting questions on how to use our probabilistic prediction of the time varying transmission strength to propagate uncertainty, in the context of variable planned interventions.

{\color{myred}
Estimation of (static) parameters is a challenging task in SMC. Several parameters related to the SEIR model, as well as parameters related to the observation processes, were pre-estimated based on external data sources. In~\citep{FHImodelPaper2021}, we used a version of ABC. Parameters in the $\{R_t\}$ dynamic process were estimated online based on the procedure of sufficient statistics, a method that can lead to degeneracy. We have also tested out the particle Metropolis-Hastings algorithm by~\citet{andrieu2010particle}. However convergence was slow and challenging, because of the computational cost of running the SMC algorithm even for a small number of particles. We validated our estimates in the supplementary material, and find that our online estimation procedure worked reasonably well for the given models and data. However, some experiments with the AR model when also including estimation of the parameter $\mu$, did cause degeneracy problems as have been reported in in~\citet{andrieu2005line}. 
A more efficient SMC algorithm might be better in utilising the potential of such algorithms to estimate static  parameters. However, our estimates of $\{R_t\}$ appear to be relatively robust with respect to changes in these parameters. Other parameters related to the SEIR model and the observation processes can be more important.}

Communicating uncertainty of estimates and the effect of stochastic and uncertain time lags from data back to infections, is a major challenge. Our current strategy has been to report estimates of the reproduction numbers one week back in time as the most reliable estimates, and this needs to be studied further. 

{\color{myred} As pointed out by one of the reviewers, it is quite easy to extend the modelling approach (and the algorithm) to include dummy variables describing interventions made by the government.
In this case, one would need to estimate a time delay between the time point in which the interventions are decided and when they effect viral transmission. This delay might change in time and be intervention specific. In our model, interventions appear in the data after a delay, and are then reflected in a change of the reproduction number. It is possible to interpret changes in the estimated reproduction numbers in the light of the interventions set in place.}

Finally, it is interesting to compare the estimates of the instantaneous reproduction number produced by this SMC model, with the estimates obtained by models which keep the reproduction number constant over longer time intervals, for example over four weeks. The SMC-based $R_t$ is able to capture changes at shorter time scales significantly better, but possibly with larger uncertainty than estimates of reproduction numbers assumed constant  over longer time periods, if the transmission has been stable during such periods. Comparing prediction power is a further aspects that can be compared.
Results from our SMC model are currently used in the weekly reports of the Norwegian Institute of Public Health, see~\href{https://www.fhi.no/en/publ/2020/weekly-reports-for-coronavirus-og-covid-19/}{www.fhi.no/en/publ/2020/weekly-reports-for-coronavirus-og-covid-19/}.

\vskip 4mm
{\color{myred}
\section*{Code and data}

The data sets analysed in this paper come from the national emergency preparedness registry for Covid-19, owned by the Norwegian Institute of Public Health. The preparedness registry is temporary and comprises data from a variety of central health registries, national clinical registries and other national administrative registries. Further information on the registry, including access to data, is available at \url{www.fhi.no/en/id/infectious-diseases/coronavirus/emergency-preparedness-register-for-covid-19/}. 

An R package called \texttt{smc.covid} is available at \href{https://github.com/geirstorvik/smc.covid}{github.com/geirstorvik/smc.covid}. It contain scripts for running the SMC algorithm on some data examples. Because data are sensitive,  not all data sets are public. The number of individuals in hospital and the number of positive cases per day are available, but the number of imported cases and the mobility matrices are confidential. For the two latter data sets we have therefore provided some simulated data sets which resemble the true ones.}

\vskip 4mm
\section*{Acknowledgement}
We acknowledge the support from many colleagues at the Norwegian Institute of Public Health in the collection and preparation of the data. Funding to this project was provided by the centre BigInsight (NFR project 237718), the project "A real-time analytical pipeline for preparedness, planning and response during the Covid-19 pandemic in Norway" (NFR project 312721) and the Nordic project "Data streams and mathematical modelling pipelines to support preparedness and decision making for Covid-19 and future pandemics" (Nordforsk NordicMathCovid project).

\vskip 4mm

\bibliographystyle{chicago}
\bibliography{references}

\appendix
\counterwithin{figure}{section}

\section{Derivations for the dynamic models for $\{R_t\}$}\label{sec:par.R}
Here we define $L_t=\log(R_t)$ for simplicity of notation and neglect the first time points (before March 7 2020) for which $R_t$ is constant. Note that for each model, the sufficient statistics involved are all easily updated with increasing $t$. 
\subsection{The random walk model}
Assume a prior $p(\tau)=\text{Gamma}(\alpha_\tau,\beta_\tau)$ with $\tau=1/\sigma_R^2$. 
Given $\{R_t\}$, the posterior for $(\phi_c,\tau)$ becomes
\begin{align*}
p(\tau|\bm R_{1:t})
\propto&\tau^{\alpha_\tau-1}\exp(-\beta_\tau\tau) \prod_{s=1}^t\tau^{1/2}\exp(-\tfrac{1}{2}\tau(L_s-L_{s-1})^2)\\
\propto&\tau^{\alpha_\tau+0.5t}\exp(-[\beta_\tau+0.5\sum_{s=1}^t(L_s-L_{s-1})^2]\tau)\\
\propto&\text{Gamma}(\tau;\alpha_\tau+0.5t,\beta_\tau+0.5\sum_{s=1}^t(L_s-L_{s-1})^2).
\end{align*}
 Simulation of $L_{t+1}$ can then be performed by first simulating 
$\tau$  from the Gamma distribution and thereafter
$R_{t+1}$ from the model.
\subsection{The piecewise constant model}
Assume a prior $p(\phi_c,\tau)=\text{Beta}(a_\phi,b_\phi)\text{Gamma}(\alpha_\tau,\beta_\tau)$ with $\tau=1/\sigma_R^2$. 
Given $\{R_t\}$ and changepoint indicators $C_t$ ($=1$ if there is a changepoint at time $t$), the posterior for $(\phi_c,\tau)$ becomes
\begin{align*}
p(\phi_c,\tau|\bm R_{1:t})
\propto&\phi_c^{a_\phi-1}(1-\phi_c)^{b_\phi-1}
        \tau^{\alpha_\tau-1}\exp(-\beta_\tau\tau)\times\\
&\prod_{s=1}^t[(1-\phi_c)^{1-C_s}\phi_c^{C_s}
       \tau^{C_s/2}\exp(-\tfrac{1}{2}\tau C_s(L_s-L_{s-1})^2)]\\
\propto&\phi_c^{a_\phi+n^c_t-1}(1-\phi_c)^{b_\phi+t-n_t^c-1}(1-\phi_c)^{n_t^c}\times\\
&\tau^{\alpha_\tau+0.5n_t^c}\exp(-[\beta_\tau+0.5\sum_{s=1}^tC_s(L_s-L_{s-1})^2]\tau)\\
\propto&\text{Beta}(\phi_c;a_\phi+n_t^c,b_\phi+t-n_t^c)\times\\
  &\text{Gamma}(\tau;\alpha_\tau+0.5n_t^c,\beta_\tau+0.5\sum_{t=1}^TC_s(L_s-L_{s-1})^2)
\end{align*}
where $n^c_t$ is the number of changepoints  up to timepoint $t$. Simulation of $R_{t+1}$ can then be performed by first simulating 
$(\phi_c,\tau)$  from the Beta-Gamma distribution and thereafter
$R_{t+1}$ from the model.

\subsection{The autoregressive model}

We consider here only the case where $\mu=0$.
Assume
\begin{align*}
p(a,\tau)=&N(a_0,[\kappa_0\tau]^{-1})\times\text{Gamma}(\alpha_\tau,\beta_\tau)
\end{align*}
Note first that
\begin{align*}
g_t(a)\equiv&\kappa_0(a-a_0)^2+\sum_{s=1}^t(L_s-aL_{s-1})^2\\
=&\kappa_0 a^2-2\kappa_0 a_0a+\kappa_0a_0^2+\sum_{s=1}^tL_s^2-2a\sum_{s=1}^tL_sL_{s-1}+a^2\sum_{s=1}^tL_{s-1}^2\\
=&(\kappa_0+\sum_{s=1}^tL_{s-1}^2)a^2-2[\kappa_0 a_0+\sum_{s=1}^tL_sL_{s-1}]a+\kappa_0 a_0^2+\sum_{s=1}^tL_s^2\\\
=&(\kappa_0+\sum_{s=1}^tL_{s-1}^2)(a-\hat a_t)^2-(\kappa_0+\sum_{s=1}^tL_{s-1}^2)\hat a_t^2+\kappa_0 a_0^2+\sum_{s=1}^tL_s^2\\
\intertext{with}
\hat a_t=&\frac{\kappa_0 a_0+\sum_{s=1}^tL_sL_{s-1}}{\kappa_0+\sum_{s=1}^tL_{s-1}^2}.
\end{align*}
This gives
\begin{align*}
p(a|\tau,R_{1:t})
\propto&\exp\left(-0.5\tau g_t(a)\right)\\
\propto& N\left(\hat a_t,\tau^{-1}[\kappa_0+\sum_{s=1}^tL_{s-1}^2]^{-1}\right).
\end{align*}
Further, for any value of $a$, we have
\begin{align*}
p(\tau|R_{1:t})
\propto&\frac{p(\tau)p(R_{1:t}|\tau,a)}{p(a|\tau,R_{1:t})}\\
\intertext{and using $a=\hat a_t(a)$, we obtain}
p(\tau|R_{1:t})\propto&\frac{\tau^{\alpha_\tau-1}\tau^{0.5t}\exp(-\beta_\tau\tau)\exp(-0.5\tau g_t(\hat a))}{\tau^{0.5}}\\
\propto&\text{ Gamma}(\alpha_\tau+0.5t,\beta_\tau+0.5g_t(\hat a_t)).
\end{align*}

\section{Parameter settings}\label{sup:parset}

The full model includes several parameter specifications that are fixed throughout all runs. Some sensitivity analysis is provided in Section~\ref{sup:add.res}.
\begin{itemize}
\item $R_t$ was set to a constant $R_0$ from February 17 2020 to March 8 2020. We estimate $R_0$ and used a $N(1.192,0.1)$ prior for $\log R_0$ (giving an expectation of $R_0$ equal to 3.3, which is compatible with the literature).
\item The amplification factor for imported cases is set to 2.8 from~\citep{FHImodelPaper2021} .
\item In order to take into account regional variability, the number of hospitalised individuals is assumed to follow a beta-binomial distribution with expectations equal to the probabilities given in Table~\ref{tab:p.h} and with a shape parameter equal to 8. The time-dynamic changes in the probabilities are based on~\citep{FHImodelPaper2021}.
\vspace*{.1cm}

\begin{table}
\begin{tabular}{c|rrrrrr}
\hline
Prob& 0.0378&0.0230&0.0209&0.0180&0.0146&0.0185\\
From day&2020-02-17&2020-05-01&2020-06-01&2020-07-01&2020-08-01&2020-09-01\\
\hline
Prob&0.0183&0.0208&0.0232&0.0221&0.0230&0.0224\\
From day&2020-10-01&2020-11-01&2020-12-01&2021-01-01&2021-02-01&2021-03-01\\
\hline
Probs&0.0231& 0.0141&0.0092\\
From day&2021-04-01& 2021-05-01&2021-06-01\\
\hline
\end{tabular}
\caption{\label{tab:p.h}Time-changing probabilities for hospitalisation, given infected. The dates correspond to starting day for the probability provided.}
\end{table}
\vspace*{.1cm}

\item The distribution for the delay from infected to hospitalised is assumed to follow a negative binomial distribution with parameters $(8.87,3.40)$ until August 1 2020 and thereafter with parameters $(7.60,3.34)$, as estimated from registry data.
\item The distribution of the time between infection and testing positive is assumed to follow a discrete distribution with probabilities $0.05,0.080,0.16,0.4,0.3,0.01$ for delays of 1-6 days. 
\item In the model~\eqref{eq:pi_test} for the detection probability, we used $\pi_0=-1.017,\pi_1=0.00013$. The number of detected individuals was assumed to follow a beta-binomial distribution with expectations given by (~\ref{eq:pi_test}) and a shape parameter equal to 8.
\end{itemize}

\section{Additional results}\label{sup:add.res}
\begin{figure}
    \centering
    \includegraphics[width=\textwidth]{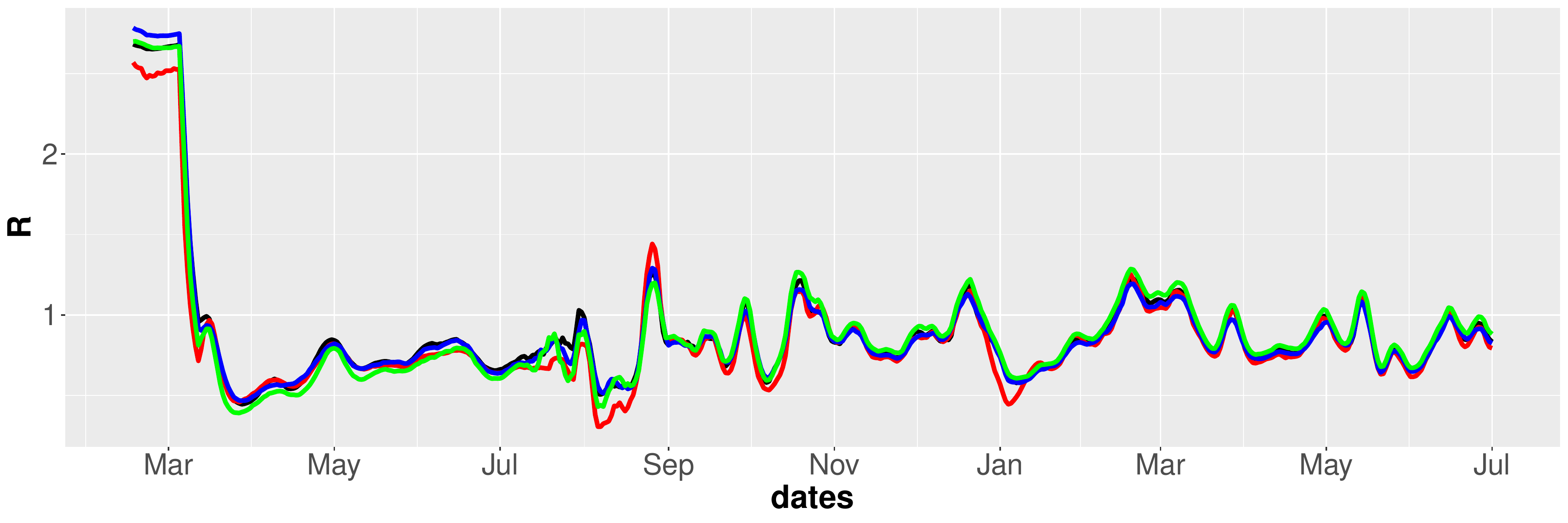}\\
    \caption{\label{fig:comp_prior}Fixed lag estimates of $R_t$ for different prior settings for the autoregressive model. Black corresponds to the settings described in section~\ref{sup:parset} and priors given in Table~\ref{tab:par.est}, the red curve is obtained by changing $(\pi_0,\pi_1)$ to (-1.175,0.000074), the green curve is obtained by changing the prior for $\tau$ to $\text{Gamma}(1.2,0.14)$ and  the blue curve corresponds to changing the $\alpha$ parameters in the Negative-binomial distributions for hospital and test data to 4.}
\end{figure}

Figure~\ref{fig:comp_prior} explores the sensitivity of results with respect to some of the parameters given in Section~\ref{sup:parset}. In this figure (where only posterior medians are shown), the black line corresponds to weekly averaged fixed lag estimates of $\{R_t\}$ based on the autoregressive model (the middle panel in Figure~\ref{fig:R_t_week}). For the red curve, we changed the prior for $\tau=1/\sigma^2$ to a $\text{Gamma}(1.2,0.14)$ distribution.
For the blue curve, we changed the values of $(\pi_0,\pi_1)$ to (-1.175,0.000074) (estimates obtained from another model on the same data). The green curve corresponds to a change in the shape parameters in the negative binomial distribution for hospitalisation and test data from 8 to 3. In all cases the results are stable, which was also the case when we compared 50\% credibility intervals (not shown).

\begin{figure}
    \centering
    \includegraphics[width=\textwidth]{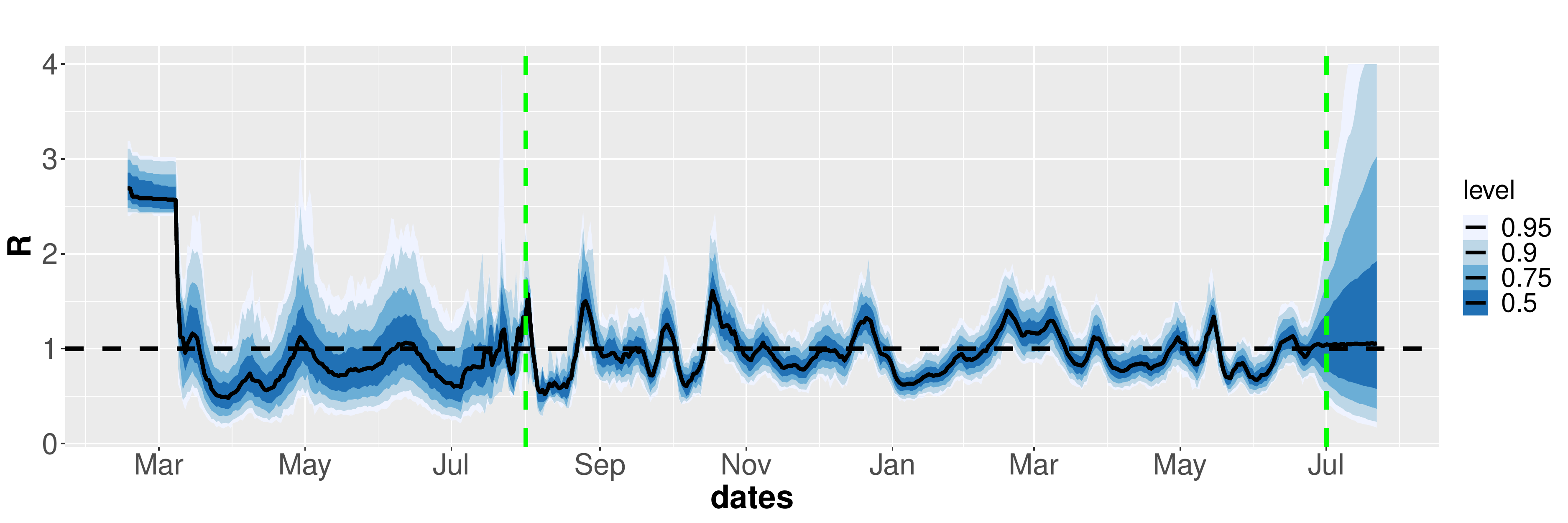}\\
    \includegraphics[width=\textwidth]{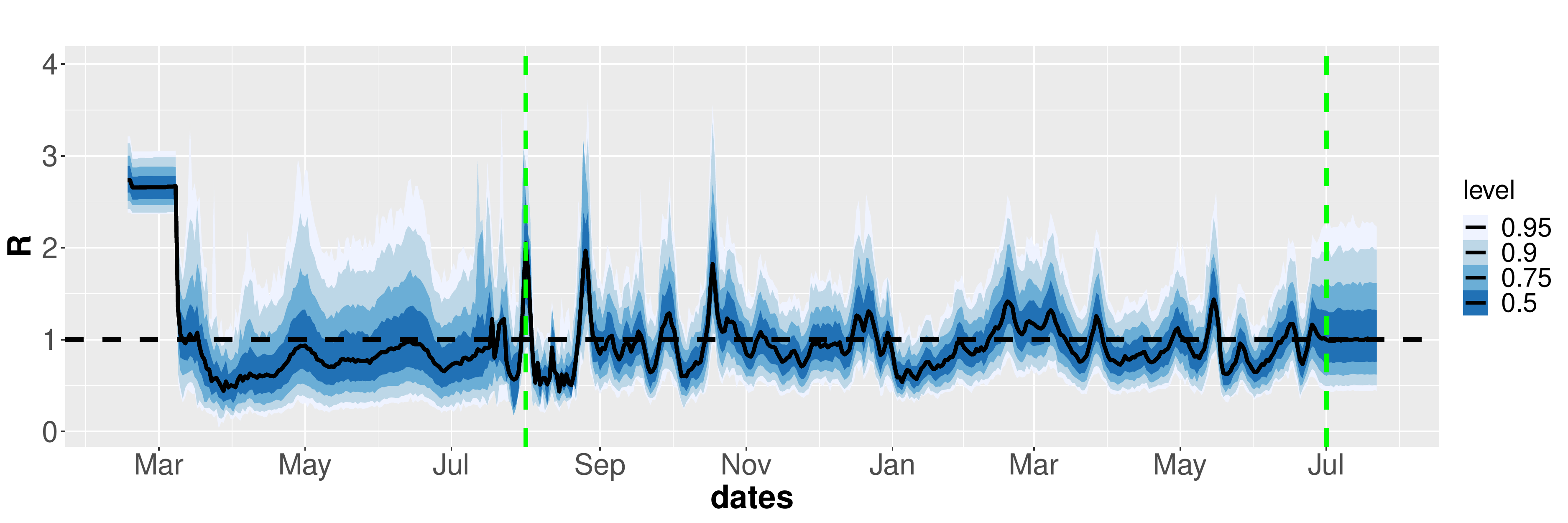}\\
    \includegraphics[width=\textwidth]{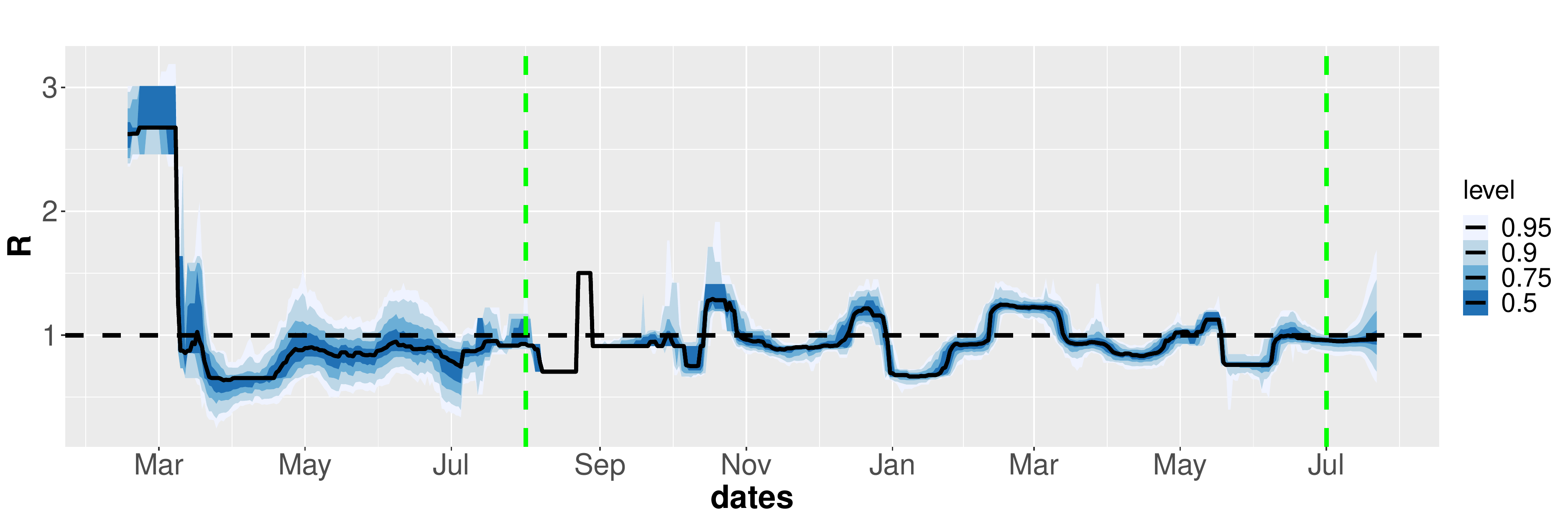}
    \caption{\label{fig:R_t_daily}Estimates of $R_t$ based on the random walk (upper), autoregressive (middle) and piece-wise constant models. Corresponding weekly averaged estimates are given in Figure~\ref{fig:R_t_week}.}
\end{figure}

Figure~\ref{fig:R_t_daily} shows the fixed-24-days-lag smoothed estimates of daily $R_t$ values for the three models. This plot is not averaged over each preceding week, as is done in  Figure~\ref{fig:R_t_week} in the main text.

\begin{figure}
    \centering
    \includegraphics[width=\textwidth]{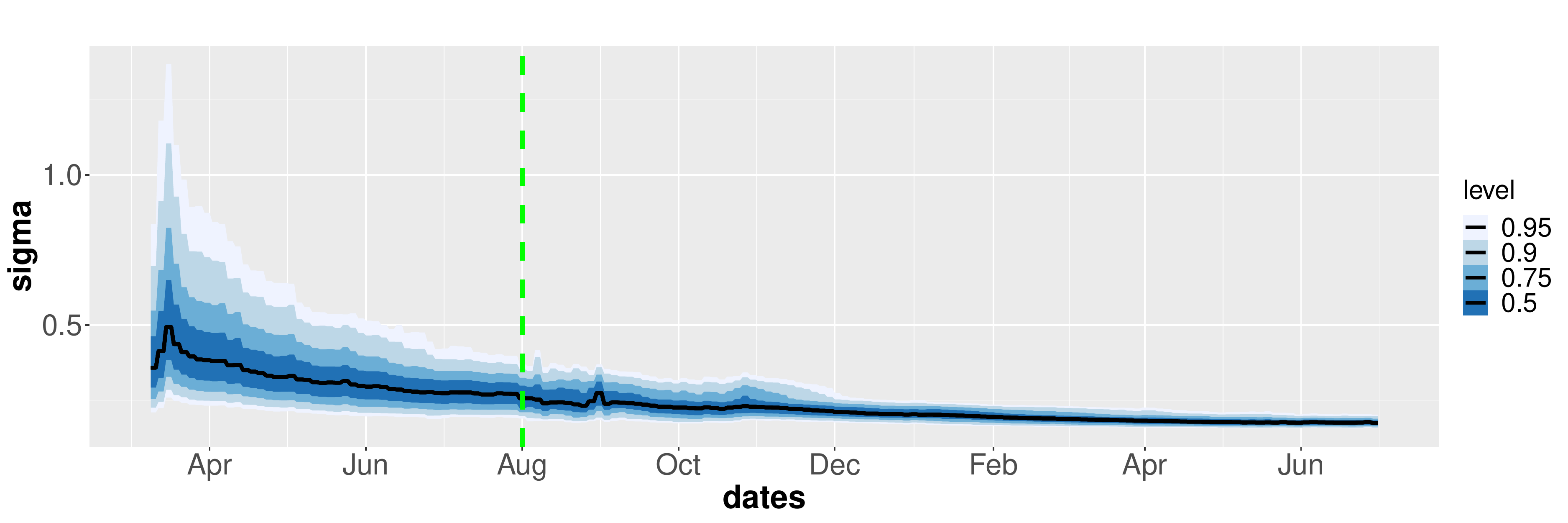}\\
    \caption{\label{fig:R_rw_par}Filtered estimates of $\sigma$ for the random walk model.}
\end{figure}

\begin{figure}
    \centering
    \includegraphics[width=\textwidth]{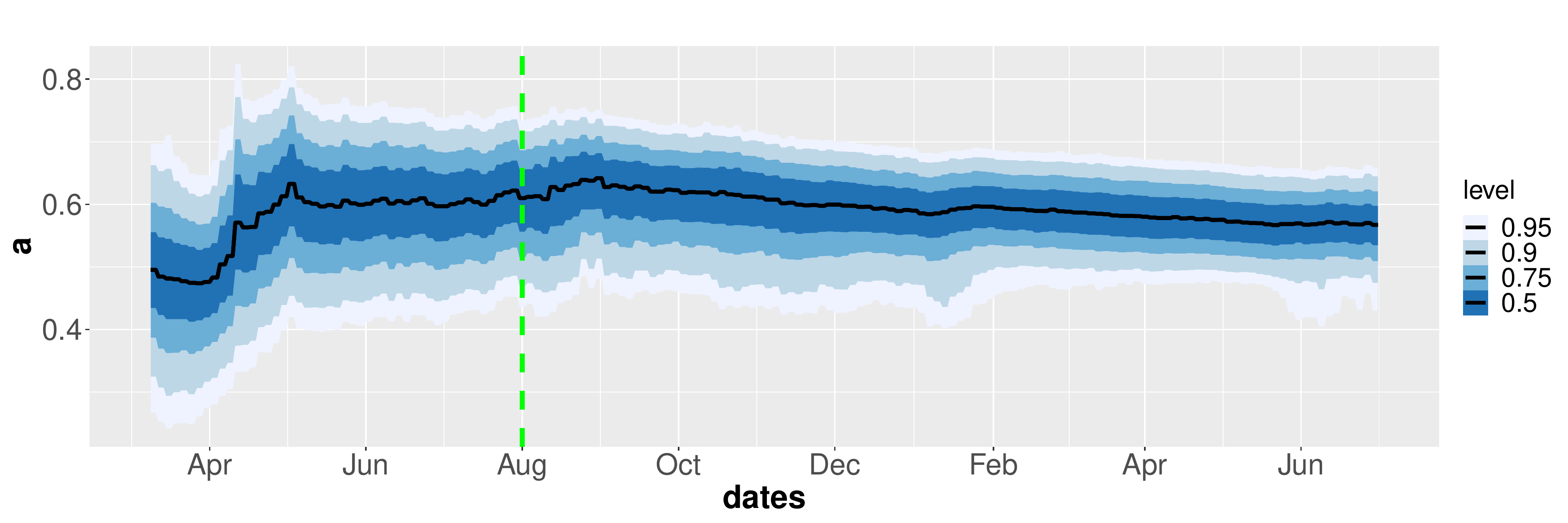}\\
    \includegraphics[width=\textwidth]{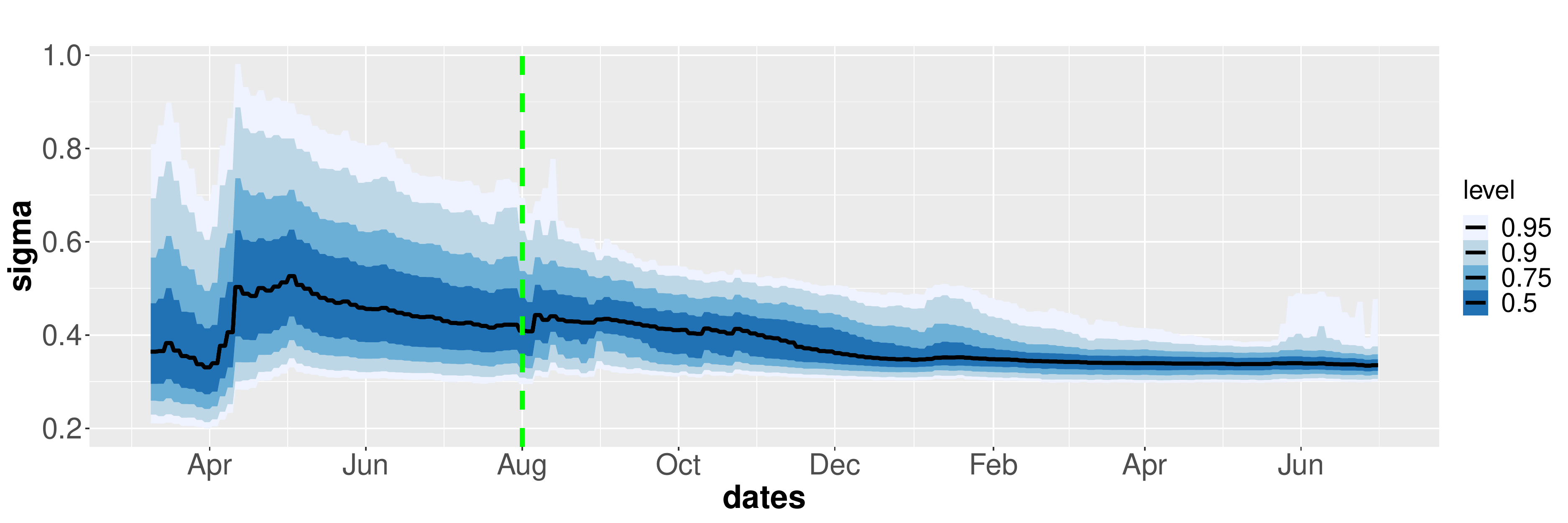}\\
    \caption{\label{fig:R_ar_par}Filtered estimates of $a$ (top) and $\sigma$ (bottom) for the autoregressive model.}
\end{figure}

\begin{figure}
    \centering
    \includegraphics[width=\textwidth]{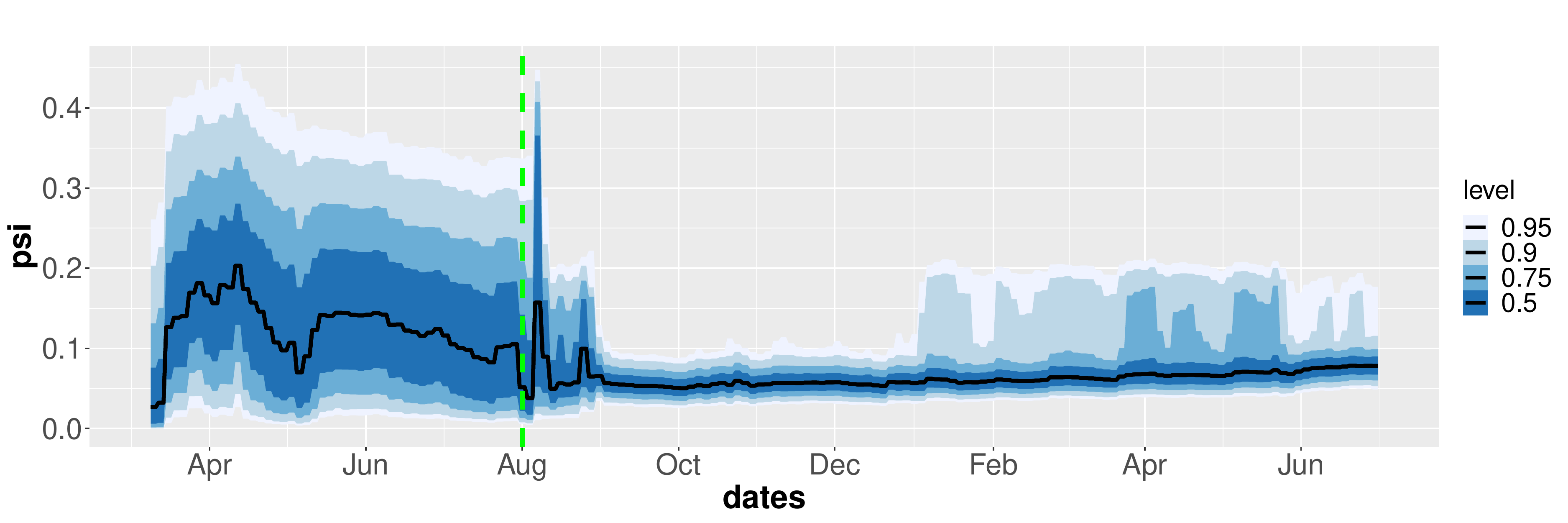}\\
    \includegraphics[width=\textwidth]{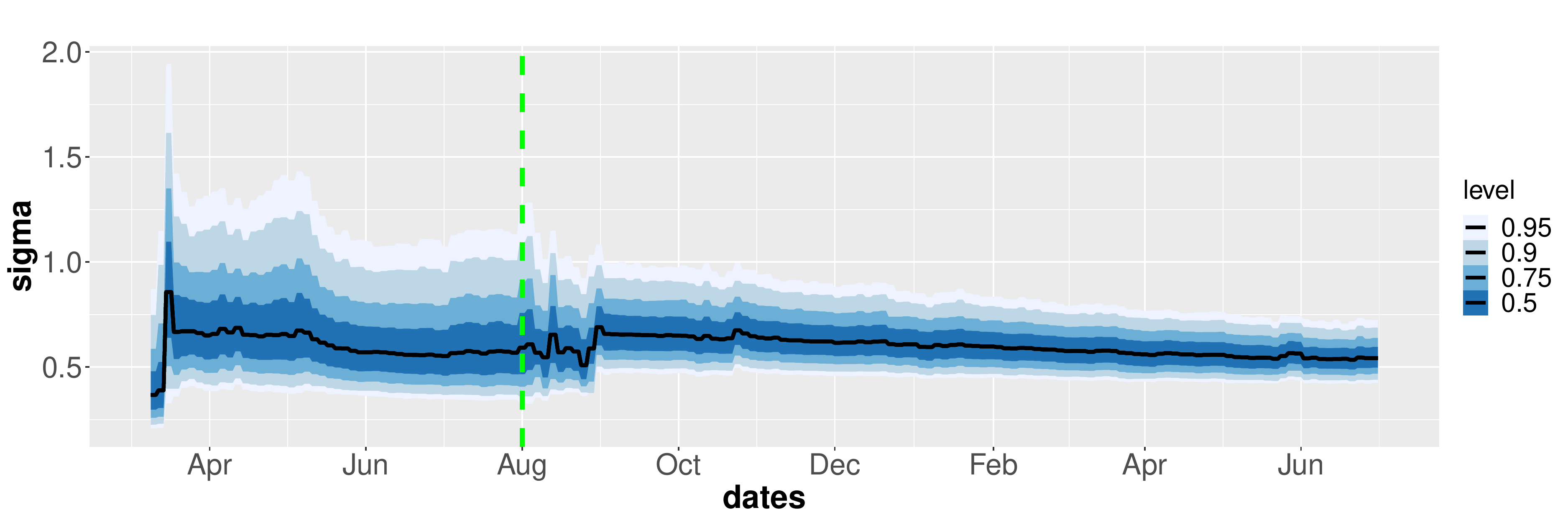}\\
    \caption{\label{fig:R_cp_par}Filtered estimates of $\phi_c$ (top) and $\sigma_R$ (bottom) for the piecewise constant model.}
\end{figure}

Figures~\ref{fig:R_rw_par}-\ref{fig:R_cp_par} show (filtered) parameter estimates of the dynamics of $\{R_t\}$ for the three models considered. All parameter estimates are based on the use of sufficient statistics~\citep{fearnhead2002markov,storvik2002particle} in the SMC algorithm. 

Table~\ref{tab:rep.mlik} shows that although there is some Monte Carlo variability in the estimation of the marginal likelihoods, the differences between the models are statistically significant. 

\begin{figure}
    \centering
    \includegraphics[width=\textwidth]{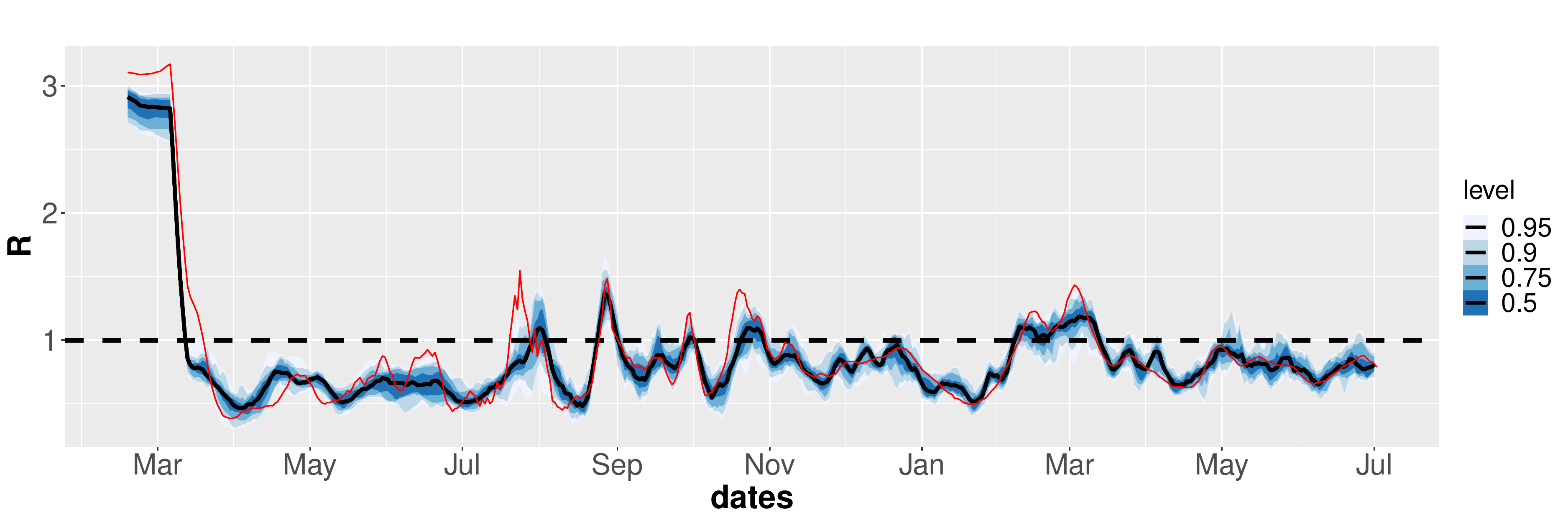}\\
    \caption{\label{fig:R_sim_rep}Results from 20 independent simulation experiments. All 20 datasets simulated based on the same true $\{R_t\}$ process (red line). The blue bands show credibility intervals of the 20 posterior medians.}
\end{figure}

{\color{myred}
Figure~\ref{fig:R_sim_rep} summarises results from the simulation experiments reported in Section~\ref{sec:sim}. For each of the 20 simulation experiments (with the same underlying true $\{R_t\}$ process), posterior medians were obtained similar as to the runs on the real data. The figure shows the variability in these medians over different data sets, demonstrating the strong information content in such data about the $\{R_t\}$ process}.

\begin{figure}
    \centering
    \includegraphics[width=\textwidth]{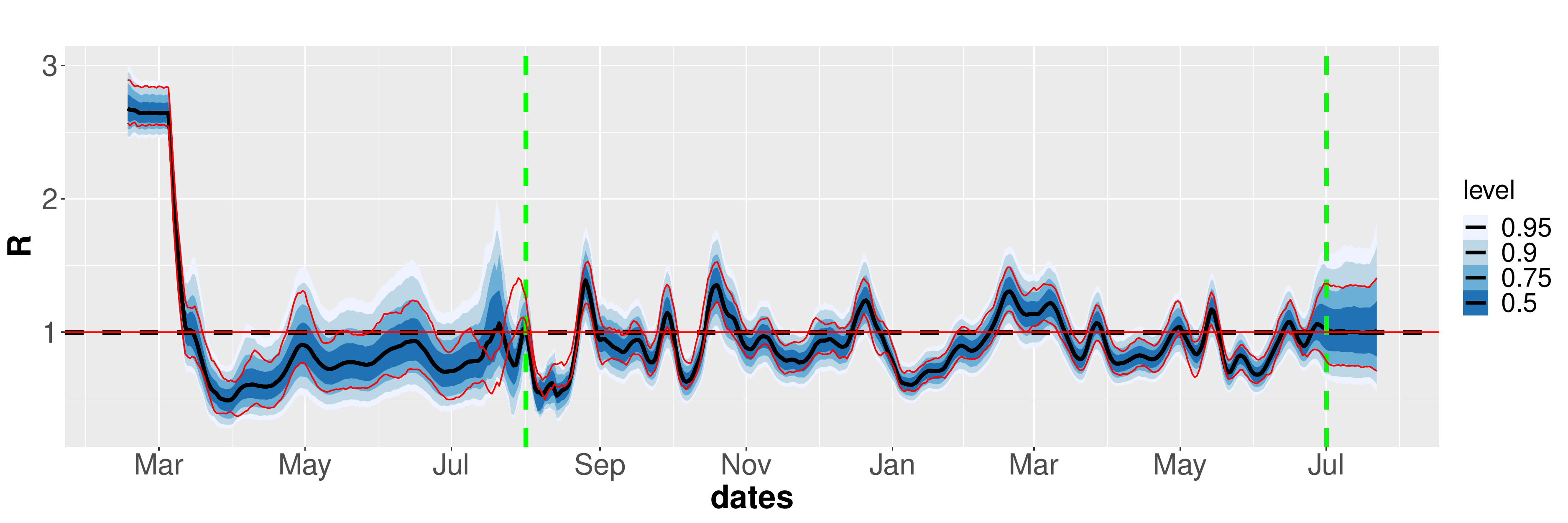}
    \caption{\label{fig:R_t_week.B}Estimates of  $\frac{1}{7} \sum_{u=0}^6R_{t-u}$  using $B=20\,000$ (blue) and $B=2\,000$ (red) particles based on the autoregressive model. Same setup as for Figure 2.}
\end{figure}

Figure~\ref{fig:R_t_week.B} gives a comparison of one run based on $B=20\,000$ particles and one run with $B=2\,000$ particles, indicating that the number of particles used are sufficient for reliable estimates of the $\{R_\}$ process.

\begin{table}[t]
\centering
\begin{tabular}{c|ccccc|c|ccc}
&\multicolumn{5}{|c|}{Repetitions}&Mean&RW&AR&CP\\
\hline
RW & -3031.56 & -3029.18 & -3033.91 & -3030.74 & -3038.64 & -3032.81 &  & 0.0000 & 0.0096 \\ 
AR & -3008.85 & -3008.06 & -3006.69 & -3010.22 & -3014.15 & -3009.60 &  &  & 0.0002 \\ 
 CP & -3026.68 & -3029.23 & -3024.41 & -3024.04 & -3018.43 & -3024.56 &  &  &  \\ 
\end{tabular}
\caption{\label{tab:rep.mlik}On the left, estimates of marginal log-likelihoods based on $B=5\,000$ particles using five repetitions of the SMC algorithm. On the right, $p$-values based on a $t$-test for testing the hypothesis on equality for the marginal log-likelihoods between the models. }
\end{table}

\vskip 4mm
{\color{myred}
\section{Estimation of hyper-parameters}\label{sec:Spar}

As described in~\citet{kantas2015particle}, there are many different approaches to estimate  static hyper-parameters, both online and offline.
In the paper we focused on online methods for estimation of parameters in the $\{R_t\}$ process, based on the use of sufficient statistics~\citep{fearnhead2002markov,storvik2002particle}. As pointed out by one of the referees, this method can suffer from degeneracy and converge towards different limit distributions in different runs. Such degeneracy problems can in particular occur for long time series. We validate the estimates obtained in two ways. We concentrate on the AR process, both due to the fact that this is our best model and that it turned out to be the most problematic one, with respect to parameter estimation. 

Figure~\ref{fig:comp_par} shows the parameter estimates obtained from two independent runs.  The agreement of these runs indicates that the estimation procedure performs quite well in our case. We do however emphasise that even though the problems with degeneracy did not seem to be present for the model considered here, if we considered an AR process where also $\mu$ is estimated, indeed different runs of the SMC algorithm resulted in somewhat different parameter estimates. The problems in that case occurred on the day where test data were introduced.

\begin{figure}
    \centering
    \includegraphics[width=\textwidth]{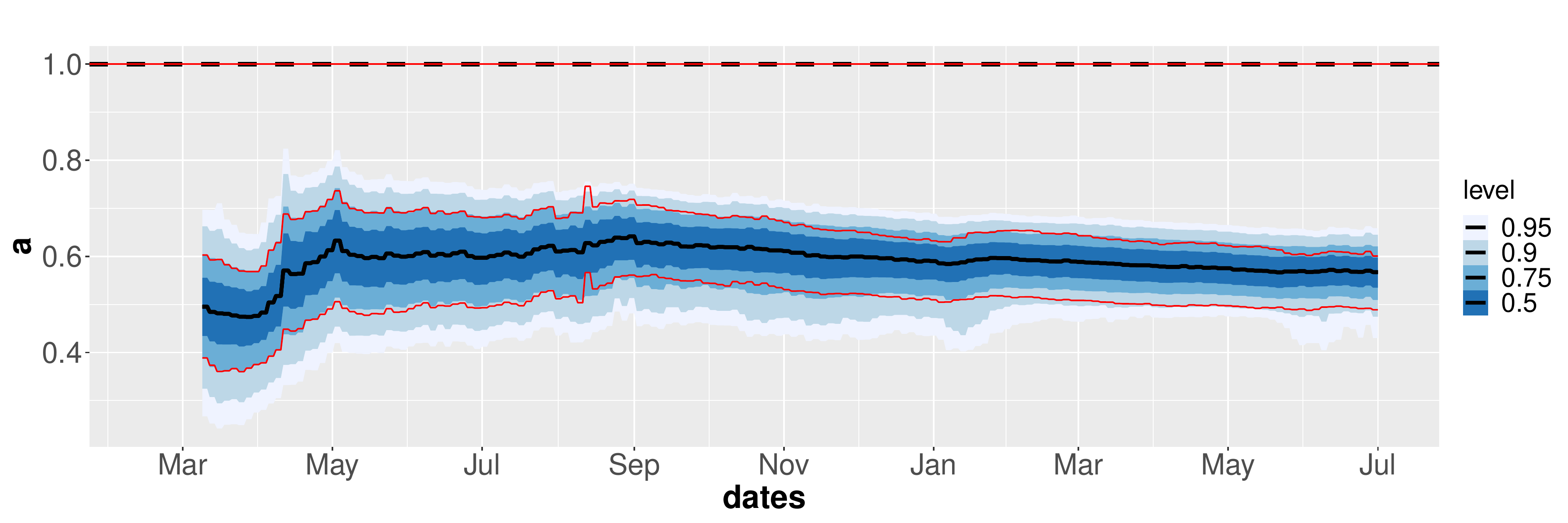}\\
    \includegraphics[width=\textwidth]{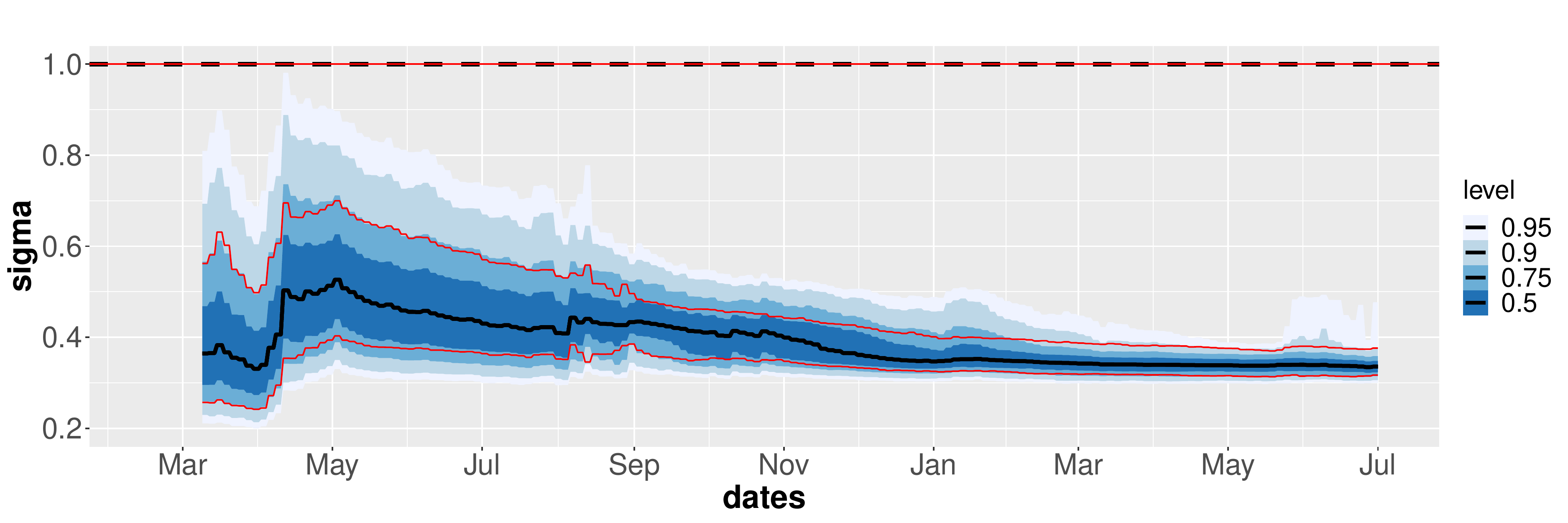}\\
    \caption{\label{fig:comp_par}Estimates of $a$ (top) and $\sigma$ (bottom) for the AR model. The blue bands show different confidence bands while the red curves show 95\% confidence bands from an independent run of the SMC algorithm. Settings are the same as for the results in Figure~\ref{fig:R_t_week}.}
\end{figure}

An important feature of SMC methods is that unbiased estimates of marginal likelihoods are available, making the pseudo-marginal approach by~\citet{andrieu2009pseudo} possible. Given a current parameter $\bm\theta$ with corresponding estimate of the marginal likelihood $\hat p(\bm y_{1:T};\bm\theta)$, a proposal $\tilde{\bm\theta}\sim q(\tilde{\bm\theta}|\bm\theta)$ is generated and an estimate $\hat p(\bm y_{1:T};\tilde{\bm\theta})$ is obtained by a new run of the SMC algorithm. The new proposal will then be accepted with probability
\[
\min\{1,\frac{p(\tilde{\bm\theta})\hat p(\bm y_{1:T};\tilde{\bm\theta})q(\bm\theta|\tilde{\bm\theta})}
{p(\bm\theta\hat) p(\bm y_{1:T};\bm\theta)q(\tilde{\bm\theta}|\bm\theta)}\},
\]
corresponding to the Particle Metropolis-Hastings Algorithm in~\citet{andrieu2010particle}.
Practical use of this algorithm requires tuning of the proposal distribution $q(\tilde{\bm\theta}|\bm\theta)$ and of the number of particles used in each SMC run.
Concerning the latter, the number of particles in each SMC run needs to be reduced dramatically compared to a single run of the SMC algorithm since these runs have to be repeated many times.  Although~\citet{andrieu2010particle} showed that the number of particles should increase with time, we have only considered a fixed number of particles in our testing. Even with many trials on the tuning parameters, we struggled to obtain convergence within reasonable time (running the algorithm for several days). However, we stored all the proposed parameter settings together with their likelihood values. We then fitted a two-dimensional GAM model to the log-likelihood values using $a$ and $\sigma$ as covariates. More than 500 samples of $(a,\sigma)$ were obtained through multiple runs  of the PMCMC algorithm with different starting points (running in total for several days). Figure~\ref{fig:par_pmcmc_gam} shows the fitted GAM curve with the red circles corresponding to the samples obtained from the SMC algorithm. The SMC samples seem to be within the higher likelihood region of the fitted curve. Note that due to lack of converge, the PMCMC algorithm has probably not been able to explore the full posterior properly. Further, the "observed" values displayed as small dots in the plot are \emph{all} proposed values, not only the accepted ones, so these points should not be used to reflect the whole distribution.

\begin{figure}
    \centering
    \includegraphics[width=\textwidth]{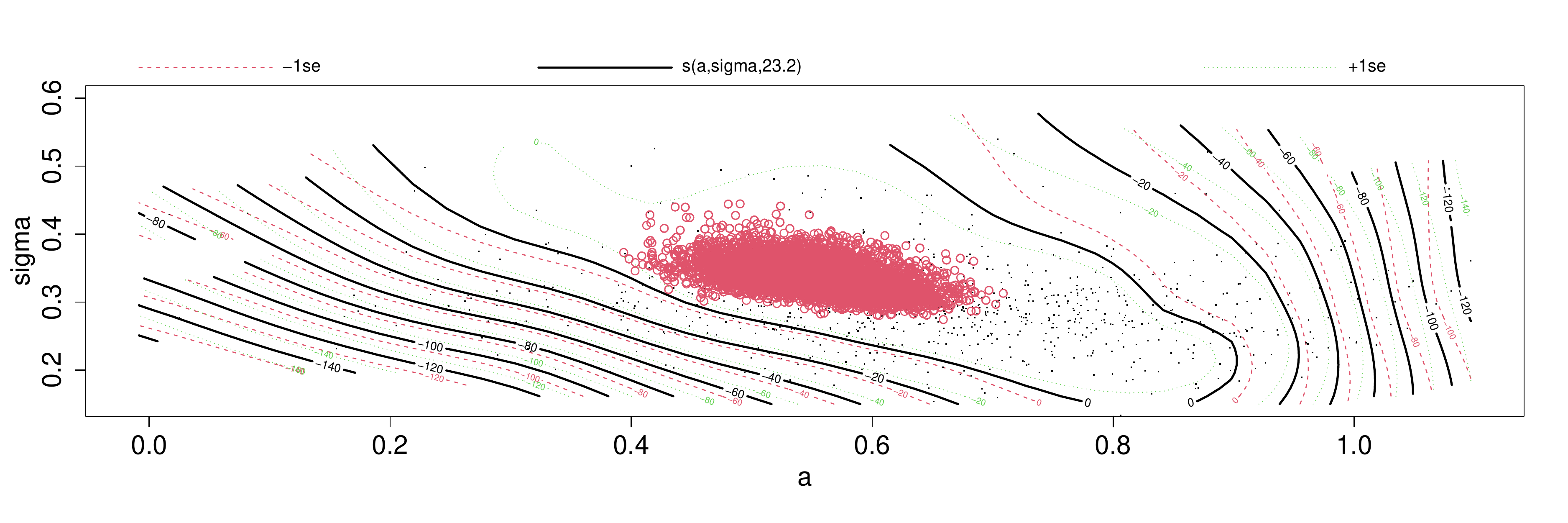}
   \caption{\label{fig:par_pmcmc_gam}Fitted two-dimensional GAM curve  estimating the log-likelihood function based on values obtained through a run of the Particle Metropolis-Hastings algorithm. $B=250$ particles were used for each run of the SMC algorithm, the proposal distributions were $N(a_{cur},1)$ for $a$ and  $N(\log\sigma_{cur},0.125)$ for $\sigma$ on log-scale. Merged results from several runs of PMCMC with different starting points.}
\end{figure}
}


\end{document}